\RequirePackage{ifpdf}
\ifpdf 
\documentclass[pdftex]{sigma}
\else
\documentclass{sigma}
\fi

\numberwithin{equation}{section}

\begin{document}

\allowdisplaybreaks

\renewcommand{\thefootnote}{$\star$}

\renewcommand{\PaperNumber}{084}

\FirstPageHeading

\ShortArticleName{Solvable Rational Potentials and Exceptional Orthogonal Polynomials}

\ArticleName{Solvable Rational Potentials\\ and Exceptional Orthogonal Polynomials\\ in Supersymmetric Quantum Mechanics\footnote{This
paper is a contribution to the Proceedings of the Eighth
International Conference ``Symmetry in Nonlinear Mathematical
Physics'' (June 21--27, 2009, Kyiv, Ukraine). The full collection
is available at
\href{http://www.emis.de/journals/SIGMA/symmetry2009.html}{http://www.emis.de/journals/SIGMA/symmetry2009.html}}}

\Author{Christiane QUESNE}

\AuthorNameForHeading{C. Quesne}

\Address{Physique Nucl\'eaire Th\'eorique et Physique Math\'ematique,  Universit\'e Libre de
Bruxelles, \\ Campus de la Plaine CP229, Boulevard~du Triomphe, B-1050 Brussels,
Belgium}
\Email{\href{mailto:cquesne@ulb.ac.be}{cquesne@ulb.ac.be}}

\ArticleDates{Received June 12, 2009, in f\/inal form August 12, 2009;  Published online August 21, 2009}

\Abstract{New exactly solvable rationally-extended radial oscillator and Scarf I potentials are generated by using a constructive supersymmetric quantum mechanical method based on a reparametrization of the corresponding conventional superpotential and on the addition of an extra rational contribution expressed in terms of some polynomial $g$. The cases where $g$ is linear or quadratic are considered. In the former, the extended potentials are strictly isospectral to the conventional ones with reparametrized couplings and are shape invariant. In the latter, there appears a variety of extended potentials, some with the same characteristics as the previous ones and others with an extra bound state below the conventional potential spectrum. Furthermore, the wavefunctions of the extended potentials are constructed. In the linear case, they contain $(\nu+1)$th-degree polynomials with $\nu=0, 1, 2, \ldots$, which are shown to be $X_1$-Laguerre or $X_1$-Jacobi exceptional orthogonal polynomials. In the quadratic case, several extensions of these polynomials appear. Among them, two dif\/ferent kinds of $(\nu+2)$th-degree Laguerre-type polynomials and a single one of $(\nu+2)$th-degree Jacobi-type polynomials with $\nu=0, 1, 2, \ldots$ are identif\/ied. They are candidates for the still unknown $X_2$-Laguerre and $X_2$-Jacobi exceptional orthogonal polynomials, respectively.}

\Keywords{Schr\"odinger equation; exactly solvable potentials; supersymmetry; orthogonal polynomials}

\Classification{33E30; 81Q05; 81Q60}

\renewcommand{\thefootnote}{\arabic{footnote}}
\setcounter{footnote}{0}

\section{Introduction}

Since the pioneering work of Bargmann \cite{bargmann}, there has been a continuing interest in generating rational potentials for which the Schr\"odinger equation is exactly or quasi-exactly solvable. In this respect, supersymmetric quantum mechanics (SUSYQM) \cite{sukumar, cooper, junker96, bagchi00, mielnik04}, as well as the related intertwining operator method \cite{infeld} and the Darboux algorithm \cite{fatveev}, have attracted considerable attention (see, e.g., \cite{mielnik84, mitra, junker97, bagchi97, blecua, gomez04a, gomez04b, carinena}).

SUSYQM indeed provides us with a powerful technique for designing Hamiltonians with a~prescribed energy spectrum from some known exactly solvable one. For one-dimensional systems, to which we are going to restrict ourselves here, the simplest approach based upon f\/irst-order intertwining operators uses a single solution of the initial Schr\"odinger equation, def\/ining the so-called superpotential. The latter then furnishes a SUSY partner Hamiltonian, whose spectrum may dif\/fer by at most one level from that of the initial Hamiltonian. More f\/lexibility may be achieved by resorting to $n$th-order intertwining operators with $n>1$, constructed from~$n$ solutions of the initial Hamiltonian \cite{andrianov95a, andrianov95b, andrianov95c, samsonov, bagchi99, aoyama, fernandez}. These higher-order operators may turn out to be reducible into a product of lower-order ones or to be irreducible.

In all cases, not only the SUSY partner potential and spectrum are explicitly known, but also the corresponding wavefunctions are given by some analytic formulae. The resulting expressions for the potential and the eigenfunctions, however, often look a good deal more complicated than those for the initial Hamiltonian (see, e.g., \cite{contreras}). In such a context, to get a SUSY partner potential that is some rational function is not a simple task in general.

An ef\/f\/icient method for getting potentials and wavefunctions expressed in terms of some elementary functions was proposed a few years ago and called algebraic Darboux transformation~\cite{gomez04a, gomez04b}. This procedure, whose idea goes back to an older paper~\cite{duistermaat}, consists in using the solutions of the hypergeometric or conf\/luent hypergeometric function whose log derivative is a~rational function and which have been classif\/ied by mathematicians \cite{erdelyi}.

In a recent work \cite{bagchi08}, an alternative scheme has been devised. Known exactly solvable potentials are rationally extended by ef\/fecting a reparametrization of the starting Hamiltonian and redef\/ining the existing superpotential in terms of modif\/ied couplings, while allowing for the presence of some rational function in it. The newly introduced parameters can be adjusted in such a way that the initial exactly solvable potential with reparametrized couplings becomes isospectral to a superposition of the same (but with its parameters left undisturbed) and some additional rational terms. Such a f\/irst-order SUSY transformation has been shown to be part of a~reducible second-order one, which explains the isospectrality of the conventional and rationally-extended potentials with the same parameters.

The purpose of the present paper is twofold: f\/irst to demonstrate the ef\/f\/iciency of this method in producing rather complicated exactly solvable rational potentials, whose existence would other\-wise have remained unknown, and second to examine in some detail the polynomials appearing in the rational potential bound-state wavefunctions. Among these polynomials, the simplest ones belong to two classes of exceptional orthogonal polynomials, which have been the topic of some recent mathematical study \cite{gomez08a, gomez08b}. We plan to show here that our SUSYQM approach provides us with a convenient way of constructing and generalizing such polynomials.

This paper is organized as follows. In Section~\ref{section2}, the method is f\/irst applied to rationally extend the radial oscillator potential and to study the associated Laguerre-type polynomials. The generalization to the Scarf I potential and to the corresponding Jacobi-type polynomials is then dealt with in Section~\ref{section3}. Finally, Section~\ref{section4} contains some final comments.

\section{Radial oscillator potential}\label{section2}

The radial oscillator potential
\begin{gather}
  V_l(x) = \frac{1}{4} \omega^2 x^2 + \frac{l(l+1)}{x^2},  \label{eq:RHO-pot}
\end{gather}
where $\omega$ and $l$ denote the oscillator frequency and the angular momentum quantum number, respectively, is def\/ined on the half-line $0 < x < \infty$. It supports an inf\/inite number of bound states, whose energies are given by
\begin{gather}
  E^{(l)}_{\nu} = \omega \left(2\nu + l + \tfrac{3}{2}\right), \qquad \nu=0, 1, 2, \ldots.  \label{eq:RHO-E}
\end{gather}
The associated wavefunctions, vanishing at the origin and decaying exponentially at inf\/inity, can be expressed in terms of Laguerre polynomials as \cite{cooper, moshinsky}
\begin{gather}
  \psi^{(l)}_{\nu}(x) = {\cal N}^{(l)}_{\nu} x^{l+1} L^{(l+\frac{1}{2})}_{\nu}(z) e^{- \frac{1}{4} \omega x^2},
  \qquad z \equiv \tfrac{1}{2} \omega x^2,  \label{eq:RHO-wf}
\end{gather}
where
\begin{gather*}
  {\cal N}^{(l)}_{\nu} = \left(\frac{\omega}{2}\right)^{\frac{1}{2} (l+\frac{3}{2})} \left(\frac{2\, \nu!}
  {\Gamma(\nu+l+\frac{3}{2})}\right)^{1/2}.
\end{gather*}

Before applying SUSYQM to such a potential, let us brief\/ly review some well-known properties of this method \cite{sukumar, cooper, junker96, bagchi00, mielnik04}.

In the minimal version of SUSY, one considers a pair of f\/irst-order dif\/ferential operators
\begin{gather}
  \hat{A} = \frac{d}{dx} + W(x), \qquad \hat{A}^{\dagger} = - \frac{d}{dx} + W(x),  \label{eq:A}
\end{gather}
and a pair of factorized Hamiltonians
\begin{gather}
  H^{(+)} = \hat{A}^{\dagger} \hat{A} = - \frac{d^2}{dx^2} + V^{(+)}(x) - E, \qquad H^{(-)} = \hat{A}
  {\hat A}^{\dagger} = - \frac{d^2}{dx^2} + V^{(-)}(x) - E,  \label{eq:H+/-}
\end{gather}
where $E$ is the so-called factorization energy. To $E$ one can associate a factorization function~$\phi(x)$, such that $H^{(+)} \phi(x) = E \phi(x)$. The superpotential is given in terms of $\phi(x)$ by the relation $W(x) = - \phi'(x)/\phi(x)$ and the two partner potentials $V^{(\pm)}(x)$ are connected with $W(x)$ through the equation
\begin{gather}
  V^{(\pm)}(x) = W^2(x) \mp W'(x) + E, \label{eq:V+/-}
\end{gather}
where a prime denotes derivative with respect to $x$.

Whenever $E$ and $\phi(x)$ are the ground-state energy $E^{(+)}_0$ and wavefunction $\psi^{(+)}_0(x)$ of the initial potential $V^{(+)}(x)$, i.e., $H^{(+)} \phi(x) = H^{(+)} \psi^{(+)}_0(x) = 0$, the partner potential $V^{(-)}(x)$ has the same bound-state spectrum as $V^{(+)}(x)$, except for the ground-state energy which is removed (case {\it i}). For $E < E^{(+)}_0$, in which case $\phi(x)$ is a nonnormalizable eigenfunction of $H^{(+)}$, the partner $V^{(-)}(x)$ has the same spectrum as $V^{(+)}(x)$ if $\phi^{-1}(x)$ is also nonnormalizable (case~{\it ii}) or has an extra bound-state energy $E$ below $E^{(+)}_0$, corresponding to the wavefunction $\phi^{-1}(x)$, if the latter is normalizable (case {\it iii}).

The standard SUSY approach to the radial oscillator potential corresponds to case {\it i}. Deletion of the ground-state energy $E^{(l)}_0$ yields a partner potential $V_{l+1}(x)$ of the same kind, this meaning that the radial oscillator potential is shape invariant~\cite{gendenshtein}. In this case, $\phi(x) = x^{l+1} \exp(- \frac{1}{4} \omega x^2)$ or, in other words,
\begin{gather}
  W(x) = \frac{1}{2} \omega x - \frac{l+1}{x}.  \label{eq:RHO-W}
\end{gather}

We now plan to modify such a superpotential in order to generate rationally-extended radial oscillator potentials.

\subsection{Extended classes of rational radial oscillator potentials}\label{section2.1}

Motivated by the form of the standard superpotential (\ref{eq:RHO-W}) and by the fact that $z = \frac{1}{2} \omega x^2$ is the basic variable appearing in the Laguerre polynomials that control the bound-state wavefunctions~(\ref{eq:RHO-wf}), let us make the ansatz
\begin{gather}
  W(x) = ax + \frac{b}{x} - \frac{dg/dx}{g},  \label{eq:ansatz}
\end{gather}
where $a$, $b$ are two constants and $g$ is some polynomial in $z$ (or simply $x^2$). Our purpose is to show that it is possible to choose the constants $a$ and $b$, as well as those appearing in $g$, in such~a~way that $V^{(+)}(x)$, as def\/ined in (\ref{eq:V+/-}), continues to belong to the radial oscillator family up to some change in $l$, while its partner $V^{(-)}(x)$ dif\/fers from $V_l(x)$ in the presence of some additional rational terms, i.e.,
\begin{gather*}
  V^{(+)}(x)  = V_{l'}(x),  \qquad
  V^{(-)}(x)  = V_{l,{\rm ext}}(x) = V_l(x) + V_{l,{\rm rat}}(x).
\end{gather*}
To be physically acceptable, such additional terms $V_{l,{\rm rat}}(x)$ must be singularity free on the half-line $0 < x < \infty$.

In the following, we carry out this programme for linear and quadratic polynomials $g(x^2)$.

\subsubsection{Linear case}\label{section2.1.1}

Let us assume that
\begin{gather*}
  g(x^2) = x^2 + c, \qquad \frac{dg(x^2)}{dx} = 2x,
\end{gather*}
where $c$ is a constant satisfying the condition $c>0$ in order to ensure the absence of poles on the half-line.

On inserting the ansatz (\ref{eq:ansatz}) in (\ref{eq:V+/-}), we obtain
\begin{gather*}
  V^{(+)}(x) = a^2 x^2 + \frac{b(b+1)}{x^2} + 2 \frac{2(ac-b)+1}{x^2 + c} + a (2b-5) + E.
\end{gather*}
Elimination of the term proportional to $(x^2 + c)^{-1}$ and of the additive constant from $V^{(+)}(x)$ can be achieved by choosing
\begin{gather*}
  c = \frac{2b-1}{2a}, \qquad E = - a (2b-5).
\end{gather*}

The partner potential is then given by
\begin{gather*}
  V^{(-)}(x) = V^{(+)} + 2W' = a^2 x^2 + \frac{b(b-1)}{x^2} + \frac{4}{x^2 + c} - \frac{8c}{(x^2 + c)^2} + 2a
\end{gather*}
and its f\/irst two terms coincide with $V_l(x)$, def\/ined in (\ref{eq:RHO-pot}), if the two conditions
\begin{gather*}
  a^2 = \tfrac{1}{4} \omega^2, \qquad b(b-1) = l(l+1)
\end{gather*}
are fulf\/illed. These two restrictions lead to four possible solutions for $(a, b)$: $(a, b) = \left(\frac{\omega}{2}, l+1\right)$, $\left(\frac{\omega}{2}, - l\right)$, $\left(- \frac{\omega}{2}, l+1\right)$, $\left(- \frac{\omega}{2}, - l\right)$.

As a result, $c$ becomes $c = \pm (2l+1)/\omega$, where only the upper sign is allowed. Since this is achieved either for $(a, b) = \left(\frac{\omega}{2}, l+1\right)$ or for $(a, b) = \left(- \frac{\omega}{2}, - l\right)$, we arrive at two distinct solutions\footnote{Strictly speaking, such a picture is only valid for positive $l$ values, since for $l=0$ we are only left with the f\/irst solution.},
\begin{gather}
  V^{(+)}(x) = V_{l\pm 1}(x),\qquad
  V^{(-)}(x) = V_l(x) + \frac{4\omega}{\omega x^2 + 2l + 1} - \frac{8\omega(2l+1)}{(\omega x^2 + 2l + 1)^2}
  \pm \omega,  \label{eq:RHO-pot-lin}
\end{gather}
corresponding to{\samepage
\begin{gather}
 W(x) = \pm \frac{1}{2} \omega x \pm \frac{l + \frac{1}{2} \pm \frac{1}{2}}{x} - \frac{2\omega x}{\omega
         x^2 + 2l + 1},  \label{eq:RHO-W-lin} \\
  E = - \omega \left(l + \tfrac{1}{2} \mp 2\right),  \label{eq:RHO-E-lin} \\
  \phi(x) = x^{\mp (l+\frac{1}{2}) - \frac{1}{2}} (\omega x^2 + 2l + 1) e^{\mp \frac{1}{4} \omega x^2},
        \label{eq:RHO-phi-lin}
\end{gather}}

\noindent
where we take either all upper or all lower signs, respectively. We can therefore go from a~conventional radial oscillator potential to an extended potential $V_{l, {\rm ext}}(x)$ in f\/irst-order SUSYQM provided we start either from $V_{l+1}(x)$ or from $V_{l-1}(x)$ and, apart from some irrelevant additive constant, the result of the extension is the same in both cases.

Since the factorization energy $E$, given in (\ref{eq:RHO-E-lin}), is smaller than the ground-state energy of $V^{(+)}(x)$, namely $E^{(l\pm1)}_0 = \omega \left(l + \frac{3}{2} \pm 1\right)$, it is clear that we must be in case {\it ii} or case {\it iii} of SUSYQM. On examining next the behaviour of the factorization function, it becomes apparent that it is the former case that applies here, namely $V^{(+)}(x)$ and $V^{(-)}(x)$ are strictly isospectral. We indeed observe that for the f\/irst choice corresponding to upper signs in (\ref{eq:RHO-phi-lin}), $\phi$ decays exponentially for $x \to \infty$, but grows as $x^{- l -1}$ for $x \to 0$, while for the second choice associated with lower signs, it vanishes for $x \to 0$, but grows exponentially for $x \to \infty$.

We conclude that the addition of the two rational terms $4\omega (\omega x^2 + 2l + 1)^{-1}$ and $- 8\omega (2l + 1) (\omega x^2 + 2l + 1)^{-2}$ to $V_l(x)$ has not changed its spectrum, which remains given by equation (\ref{eq:RHO-E}).

\subsubsection{Quadratic case}\label{section2.1.2}

For
\begin{gather*}
  g(x^2) = x^4 + cx^2 + d, \qquad \frac{dg(x^2)}{dx} = 2x (2x^2 + c),
\end{gather*}
where $c$ and $d$ are two constants, the picture gets more involved because the singularity-free character of the extended potential on the half-line can be ensured in two dif\/ferent ways: the quadratic polynomial $g(x^2)$ may have either two real and negative roots or two complex conjugate ones. In association with these two possibilities, we will have to distinguish between the two cases
\begin{gather}
  c > 0, \qquad 0 < d \le \frac{c^2}{4},  \label{eq:RHO-C1}
\end{gather}
and
\begin{gather}
  d > \frac{c^2}{4}.  \label{eq:RHO-C2}
\end{gather}

On proceeding as in the linear case, it is straightforward to get
\begin{gather*}
  V^{(+)}(x) = a^2 x^2 + \frac{b(b+1)}{x^2} + 2 \frac{2(ac-2b+3)x^2 + 4ad - (2b-1)c}{x^4 + cx^2 + d} + a(2b-9)
  + E,
\end{gather*}
leading to the conditions
\begin{gather*}
  c = \frac{2b-3}{a}, \qquad d = \frac{(2b-1)(2b-3)}{4a^2}, \qquad E = - a(2b-9).
\end{gather*}
Furthermore
\begin{gather*}
  V^{(-)}(x) = a^2 x^2 + \frac{b(b-1)}{x^2} + \frac{4(2x^2-c)}{x^4 + cx^2 + d} + \frac{8(c^2-4d)x^2}{(x^4
  + cx^2 + d)^2} + 2a
\end{gather*}
yields the same four possibilities for $(a,b)$ as before. To go further, we have to distinguish between (\ref{eq:RHO-C1}) and (\ref{eq:RHO-C2}).

In the real case, the condition $c>0$ can be achieved by imposing $a = \frac{1}{2} \omega$, $b = l+1$ with $l>0$, or $a = - \frac{1}{2} \omega$, $b = 1$, or else $a = - \frac{1}{2} \omega$, $b = - l$. The next restriction $d > 0$, being equivalent to $(2b-1)/a > 0$, discards the choice $a = - \frac{1}{2} \omega$, $b = 1$. Finally, $d \le \frac{c^2}{4}$ turns out to impose $a < 0$. Hence we are left with a single solution corresponding to
\begin{gather*}
  a = - \frac{1}{2} \omega, \qquad b = - l, \qquad c = \frac{2\gamma}{\omega}, \qquad d = \frac{\gamma
  (\gamma-2)}{\omega^2}, \qquad \gamma \equiv 2l + 3.
\end{gather*}
As a result, we get for $l>0$ the pair of partner potentials
\begin{gather}
  V^{(+)}(x) = V_{l-1}(x),\nonumber
\\
  V^{(-)}(x) = V_l(x) + \frac{8\omega (\omega x^2-\gamma)}{(\omega x^2+\gamma)^2 - 2\gamma} +
  \frac{64\gamma \omega^2 x^2}{[(\omega x^2+\gamma)^2 - 2\gamma]^2} - \omega, \qquad \gamma \equiv
   2l + 3,  \label{eq:RHO-pot-I}
\end{gather}
corresponding to
\begin{gather}
  W(x) = - \frac{1}{2} \omega x - \frac{l}{x} - \frac{4\omega x (\omega x^2+\gamma)}{(\omega x^2+\gamma)
  ^2 - 2\gamma},
\nonumber\\
  E = - \omega \left(l + \tfrac{9}{2}\right),
\nonumber\\
  \phi(x) = x^l [(\omega x^2+\gamma)^2 - 2\gamma] e^{\frac{1}{4} \omega x^2}.  \label{eq:RHO-phi-I}
\end{gather}

Turning ourselves to the complex case, we observe that the single condition $d > \frac{c^2}{4}$ is equivalent to $b > \frac{3}{2}$, which imposes $b = l+1$ with $l>0$. The sign of $a$ remaining undetermined, we have actually here two distinct solutions according to the choice made, namely
\begin{gather*}
  a = \pm \frac{1}{2} \omega, \qquad b = l + 1 \quad (l>0), \qquad c = \pm \frac{2\gamma}{\omega}, \qquad
  d = \frac{\gamma (\gamma+2)}{\omega^2}, \qquad \gamma \equiv 2l - 1.
\end{gather*}
They lead to
\begin{gather}
V^{(+)}(x) = V_{l+1}(x),
\nonumber\\
  V^{(-)}(x) = V_l(x) + \frac{8\omega (\omega x^2\mp\gamma)}{(\omega x^2\pm\gamma)^2 + 2\gamma} -
  \frac{64\gamma \omega^2 x^2}{[(\omega x^2\pm\gamma)^2 + 2\gamma]^2} \pm \omega, \qquad
  \gamma \equiv 2l - 1,  \label{eq:RHO-pot-II-III}
\end{gather}
for $l>0$. Correspondingly
\begin{gather}
  W(x) = \pm \frac{1}{2} \omega x + \frac{l+1}{x} - \frac{4\omega x (\omega x^2\pm\gamma)}{(\omega x^2
  \pm\gamma)^2 + 2\gamma},
\nonumber\\
  E = \mp \omega \left(l - \tfrac{7}{2}\right),
\nonumber\\
  \phi(x) = x^{-l-1} [(\omega x^2\pm\gamma)^2 + 2\gamma] e^{\mp\frac{1}{4} \omega x^2}.
  \label{eq:RHO-phi-II-III}
\end{gather}

In comparison with the linear case which has provided us with a single rationally-extended potential, the quadratic one reveals itself much more f\/lexible since three distinct rationally-extended potentials with $l>0$ have been obtained. It is also worth observing that the angular momentum $l'$ of the starting potential has now a strong inf\/luence on the resulting extended one.

Furthermore, we note that in all three cases, the factorization energy is smaller than the ground-state energy of $V^{(+)}(x)$, as in the linear case. As far as the factorization function is concerned, however, a distinction must be made between the function in (\ref{eq:RHO-phi-I}) or that with upper signs in (\ref{eq:RHO-phi-II-III}), on one hand, and that with lower signs in the latter equation, on the other hand. For the f\/irst two functions, the situation is similar to the previous one leading to strictly isospectral partner potentials (case {\it ii}), whereas, for the third function, we observe that $\phi^{-1}(x)$ is normalizable on the half-line and is therefore the ground-state wavefunction of the corresponding $V^{(-)}(x)$ potential with an energy eigenvalue $E^{(-)}_0 = E^{(l+1)}_0 - 6\omega = \omega \left(l - \frac{7}{2}\right)$.

In the following, we shall refer to the potentials (\ref{eq:RHO-pot-I}) and (\ref{eq:RHO-pot-II-III}) with upper/lower signs as case I, II, and III potentials, respectively.

\subsection{Determination of wavefunctions}\label{section2.2}

The factorized forms (\ref{eq:H+/-}) of $H^{(+)}$ and $H^{(-)}$ imply the existence of the intertwining rela\-tions \cite{sukumar, cooper, junker96, bagchi00, mielnik04}
\begin{gather*}
  H^{(+)} \hat{A}^{\dagger} = \hat{A}^{\dagger} H^{(-)}, \qquad \hat{A} H^{(+)} = H^{(-)} \hat{A},
\end{gather*}
which enable us to determine the wavefunctions of $H^{(-)}$ from those of $H^{(+)}$ with the same energy (or vice versa). Hence, on starting from $H^{(+)} \psi^{(+)}_{\nu}(x) = \varepsilon_{\nu} \psi^{(+)}_{\nu}(x)$ with $\varepsilon_{\nu} = E^{(l')}_{\nu} - E$, we can write
\begin{gather}
  \psi^{(-)}_{\nu}(x) = \frac{1}{\sqrt{\varepsilon_{\nu}}} \hat{A} \psi^{(+)}_{\nu}(x), \qquad \nu = 0, 1, 2, \ldots,
  \label{eq:wf-ii}
\end{gather}
or
\begin{gather}
  \psi^{(-)}_0(x) \propto \phi^{-1}(x),
\qquad
  \psi^{(-)}_{\nu+1}(x) = \frac{1}{\sqrt{\varepsilon_{\nu}}} \hat{A} \psi^{(+)}_{\nu}(x), \qquad \nu = 0, 1, 2,
  \ldots,  \label{eq:wf-iii}
\end{gather}
according to whether  case {\it ii} or case {\it iii} of SUSYQM applies.

We will now proceed to use equations (\ref{eq:wf-ii}) and (\ref{eq:wf-iii}) to determine the wavefunctions of the rationally-extended radial oscillator potentials constructed in Section~\ref{section2.1}.

\subsubsection{Linear case}

In the linear case, the result for $\psi^{(-)}_{\nu}(x)$ should be the same whether we start from $\psi^{(+)}_{\nu}(x) = \psi^{(l+1)}_{\nu}(x)$ or from $\psi^{(+)}_{\nu}(x) = \psi^{(l-1)}_{\nu}(x)$. Considering the latter case to start with, we can rewrite the operator $\hat{A}$ as
\begin{gather*}
  \hat{A} = \sqrt{2\omega z} \left(\frac{d}{dz} - \frac{1}{2} - \frac{l}{2z} - \frac{2}{2z + 2l + 1}\right), \qquad
  z \equiv \frac{1}{2} \omega x^2.
\end{gather*}
Here we have used equation (\ref{eq:A}) with $W(x)$ given in (\ref{eq:RHO-W-lin}), where we assume the lower sign choice. On applying this operator on the right-hand side of equation (\ref{eq:RHO-wf}) with $l-1$ substituted for $l$, and employing equation (\ref{eq:wf-ii}), we get
\begin{gather}
  \psi^{(-)}_{\nu}(x) = \frac{2\omega}{\sqrt{\varepsilon_{\nu}}} {\cal N}^{(+)}_{\nu} \frac{x^{l+1} e^{-
  \frac{1}{4} \omega x^2}}{\omega x^2 + 2l + 1} \hat{\cal O}^{(\alpha)}_1 L^{(\alpha-1)}_{\nu}(z),
  \label{eq:RHO-partner-wf}
\end{gather}
where ${\cal N}^{(+)}_{\nu} = {\cal N}^{(l-1)}_{\nu}$, $\varepsilon_{\nu} = 2\omega\left(\nu + l + \frac{3}{2}
\right)$, $\alpha = l + \frac{1}{2}$, and we have def\/ined
\begin{gather*}
  \hat{\cal O}^{(\alpha)}_1 \equiv (z + \alpha)\left(\frac{d}{dz} - 1\right) - 1.
\end{gather*}

The action of the f\/irst-order dif\/ferential operator $\hat{\cal O}^{(\alpha)}_1$ on the Laguerre polynomial $L^{(\alpha-1)}_{\nu}(z)$ can be inferred from known dif\/ferential and recursion relations of the latter \cite{abramowitz}. The result can be written as
\begin{gather*}
  \hat{\cal O}^{(\alpha)}_1 L^{(\alpha-1)}_{\nu}(z) = \hat{L}^{(\alpha)}_{\nu+1}(z),
\end{gather*}
where $\hat{L}^{(\alpha)}_{\nu+1}(z)$ is a $(\nu+1)$th-degree polynomial, def\/ined by{\samepage
\begin{gather}
  \hat{L}^{(\alpha)}_{\nu+1}(z) = - (z + \alpha + 1) L^{(\alpha)}_{\nu}(z) + L^{(\alpha)}_{\nu-1}(z)
  \label{eq:L-lin-def}
\end{gather}
in terms some Laguerre ones.}

The partner potential wavefunctions (\ref{eq:RHO-partner-wf}) can therefore be expressed in terms of
$\hat{L}^{(\alpha)}_{\nu+1}(z)$ as
\begin{gather}
  \psi^{(-)}_{\nu}(x) = {\cal N}^{(-)}_{\nu} \frac{x^{l+1}}{\omega x^2 + 2l + 1} \hat{L}^{\left(l+\frac{1}{2}
  \right)}_{\nu+1}\left(\tfrac{1}{2}\omega x^2\right) e^{- \frac{1}{4} \omega x^2},  \label{eq:RHO-partner-wf-bis}
\end{gather}
where
\begin{gather*}
  {\cal N}^{(-)}_{\nu} = \left(\frac{\omega^{l+\frac{3}{2}} \nu!}{2^{l-\frac{3}{2}} \left(\nu + l + \frac{3}{2}
  \right) \Gamma\left(\nu + l + \frac{1}{2}\right)}\right)^{1/2}.
\end{gather*}

Similarly, for the other choice $\psi^{(+)}_{\nu}(x) = \psi^{(l+1)}_{\nu}(x)$, we can write
\begin{gather*}
  \psi^{(-)}_{\nu}(x) = \frac{4}{\sqrt{\varepsilon_{\nu}}} {\cal N}^{(+)}_{\nu} \frac{x^{l+1} e^{-
  \frac{1}{4} \omega x^2}}{\omega x^2 + 2l + 1} \hat{\cal O}^{(\alpha)}_2 L^{(\alpha+1)}_{\nu}(z),
\end{gather*}
where ${\cal N}^{(+)}_{\nu} = {\cal N}^{(l+1)}_{\nu}$, $\varepsilon_{\nu} = 2\omega\left(\nu + l + \frac{1}{2}
\right)$, $\alpha = l + \frac{1}{2}$, and
\begin{gather*}
  \hat{\cal O}^{(\alpha)}_2 \equiv (z + \alpha)\left(z \frac{d}{dz} + \alpha + 1\right) - z.
\end{gather*}
On taking into account that this time
\begin{gather*}
  \hat{\cal O}^{(\alpha)}_2 L^{(\alpha+1)}_{\nu}(z) = - (\nu + \alpha) \hat{L}^{(\alpha)}_{\nu+1}(z),
\end{gather*}
we arrive (up to some irrelevant overall sign) at the same wavefunctions (\ref{eq:RHO-partner-wf-bis}) as in the previous derivation, as it should be.

At this stage, some observations are in order. In~\cite{gomez08a, gomez08b}, it has been shown that the family of ($\nu+1$)th-degree polynomials $\hat{L}^{(\alpha)}_{\nu+1}(z)$, $\nu=0, 1, 2,\ldots$, def\/ined in equation (\ref{eq:L-lin-def}), arises in a natural way when extending Bochner's theorem on the relations between classical orthogonal polynomials and solutions of second-order eigenvalue equations \cite{bochner} by dropping the assumption that the f\/irst element of the orthogonal polynomial sequence be a constant. It is actually one of the two complete orthogonal sets of polynomials with respect to some positive-def\/inite measure that can be constructed by starting instead with some linear polynomial (the second one being considered in Section~\ref{section3.2}). For this reason, the exceptional polynomials $\hat{L}^{(\alpha)}_{\nu+1}(z)$ have been called $X_1$-Laguerre polynomials \cite{gomez08a, gomez08b}. They satisfy a second-order eigenvalue equation with rational coef\/f\/icients and are normalized in such a way that their highest-degree term is equal to $(-1)^{\nu+1} z^{\nu+1}/\nu!$ (as compared with $(-1)^{\nu} z^{\nu}/\nu!$ for $L^{(\alpha)}_{\nu}(z)$). On combining their def\/inition~(\ref{eq:L-lin-def}) with the recursion relation of classical Laguerre polynomials, one can write them as linear combinations of three Laguerre polynomials with constant coef\/f\/icients,
\begin{gather}
  \hat{L}^{\alpha}_{\nu+1}(z) = (\nu+1) L^{\alpha}_{\nu+1}(z) - 2(\nu+\alpha+1) L^{\alpha}_{\nu}(z) +
  (\nu+\alpha+1) L^{\alpha}_{\nu-1}(z).  \label{eq:L-lin-comb}
\end{gather}
A f\/irst few polynomials are listed in Appendix~\ref{appendixA}, where they may be contrasted with the corresponding classical Laguerre ones.

The extended radial oscillator potential~(\ref{eq:RHO-pot-lin}) and its corresponding wavefunctions~(\ref{eq:RHO-partner-wf-bis}) were derived for the f\/irst time in~\cite{cq08} by using the point canonical transformation method \cite{bhatta} and the dif\/ferential equation satisf\/ied by $\hat{L}^{(\alpha)}_{\nu+1}(z)$. In the same work, the potential was also shown to be shape invariant with a partner given by $V_{l+1,{\rm ext}}(x)$. Here we have adopted another type of approach based on f\/irst-order SUSYQM, which has the advantage of def\/ining the exceptional polynomials through the action of some f\/irst-order dif\/ferential operator on a classical Laguerre polynomial. As we now plan to show, this procedure can be easily extended to the quadratic case for which it will lead to some novel results.

\subsubsection{Quadratic case}\label{section2.2.2}

As proved in Section~\ref{section2.1.2}, the quadratic case leads to three distinct new rationally-extended potentials, of which the f\/irst two are similar to that obtained in the linear case (case {\it ii} of SUSYQM), while the third one is not (case {\it iii} of SUSYQM).

By proceeding as in the linear case and def\/ining $z = \frac{1}{2} \omega x^2$ and $\alpha = l + \frac{1}{2}$, we arrive at the following three f\/irst-order dif\/ferential operators
\begin{gather*}
  \tilde{\cal O}^{(\alpha)}_1 \equiv [z^2 + 2(\alpha+1)z + \alpha(\alpha+1)] \left(\frac{d}{dz} - 1\right)
  - 2(z+\alpha+1),
\\
  \tilde{\cal O}^{(\alpha)}_2 \equiv [z^2 + 2(\alpha-1)z + \alpha(\alpha-1)] \left(z \frac{d}{dz} + \alpha + 1\right)
  - 2z(z+\alpha-1),
\\
  \tilde{\cal O}^{(\alpha)}_3 \equiv [z^2 - 2(\alpha-1)z + \alpha(\alpha-1)] \left(z \frac{d}{dz} - z + \alpha
  +1\right) - 2z(z-\alpha+1),
\end{gather*}
leading to three distinct families of Laguerre-type polynomials,
\begin{gather}
  \tilde{\cal O}^{(\alpha)}_1 L^{(\alpha-1)}_{\nu}(z)  = - \tilde{L}^{(\alpha)}_{1,\nu+2}(z) \nonumber\\
\phantom{\tilde{\cal O}^{(\alpha)}_1 L^{(\alpha-1)}_{\nu}(z) }{}  = - [z^2 + 2(\alpha+2)z + (\alpha+1)(\alpha+2)] L^{(\alpha)}_{\nu}(z) + 2(z + \alpha+1)
       L^{(\alpha)}_{\nu-1}(z),
  \label{eq:L-quad-def1}
\\
    \tilde{\cal O}^{(\alpha)}_2 L^{(\alpha+1)}_{\nu}(z)  =  (\nu+\alpha-1) \tilde{L}^{(\alpha)}_{2,\nu+2}(z)\nonumber \\
\phantom{\tilde{\cal O}^{(\alpha)}_2 L^{(\alpha+1)}_{\nu}(z)}{}
= \{(\nu+\alpha-1) [z^2 + 2\alpha z + (\alpha-1)(\alpha+2)] + 2(\alpha-1)\} L^{(\alpha)}_{\nu}(z) \nonumber \\
\phantom{\tilde{\cal O}^{(\alpha)}_2 L^{(\alpha+1)}_{\nu}(z)=}{}- 2(\nu+\alpha) (z + \alpha-1) L^{(\alpha)}_{\nu-1}(z),
 \label{eq:L-quad-def2}
\\
    \tilde{\cal O}^{(\alpha)}_3 L^{(\alpha+1)}_{\nu}(z)  =  (\nu+3) \tilde{L}^{(\alpha)}_{3,\nu+3}(z) \nonumber\\
\phantom{\tilde{\cal O}^{(\alpha)}_3 L^{(\alpha+1)}_{\nu}(z)}{}
= \{- z^3 + (2\nu+3\alpha-3)z^2 - [2(2\alpha-3)\nu + 3\alpha(\alpha-1)]z \nonumber\\
\phantom{\tilde{\cal O}^{(\alpha)}_3 L^{(\alpha+1)}_{\nu}(z)=}{}
+ (\alpha-1)[2(\alpha-1)\nu +\alpha(\alpha+1)]\} L^{(\alpha)}_{\nu}(z) \nonumber\\
\phantom{\tilde{\cal O}^{(\alpha)}_3 L^{(\alpha+1)}_{\nu}(z)=}{}  - (\nu+\alpha)[z^2 - 2(\alpha-2)z + (\alpha-1)(\alpha-2)] L^{(\alpha)}_{\nu-1}(z),
 \label{eq:L-quad-def3}
\end{gather}
for cases I, II, and III, respectively.

The corresponding rationally-extended potential wavefunctions can be written as
\begin{gather*}
  \psi^{(-)}_{\nu}(x) = {\cal N}^{(-)}_{\nu} \frac{x^{l+1}}{(\omega x^2 + 2l + 3)^2 - 2(2l+3)}
  \tilde{L}^{\left(l+\frac{1}{2}\right)}_{1,\nu+2}\left(\tfrac{1}{2}\omega x^2\right) e^{- \frac{1}{4} \omega x^2}
  \qquad \mbox{\rm (case I)},
\\
  \psi^{(-)}_{\nu}(x) = {\cal N}^{(-)}_{\nu} \frac{x^{l+1}}{(\omega x^2 + 2l - 1)^2 + 2(2l-1)}
  \tilde{L}^{\left(l+\frac{1}{2}\right)}_{2,\nu+2}\left(\tfrac{1}{2}\omega x^2\right) e^{- \frac{1}{4} \omega x^2}
  \qquad \mbox{\rm (case II)},
\\
  \psi^{(-)}_{\nu+1}(x) = {\cal N}^{(-)}_{\nu+1} \frac{x^{l+1}}{(\omega x^2 - 2l + 1)^2 + 2(2l-1)}
  \tilde{L}^{\left(l+\frac{1}{2}\right)}_{3,\nu+3}\left(\tfrac{1}{2}\omega x^2\right) e^{- \frac{1}{4} \omega x^2}
  \qquad \mbox{\rm (case III)},
\end{gather*}
where{\samepage
\begin{gather*}
  {\cal N}^{(-)}_{\nu} = - \left(\frac{\omega^{l+\frac{3}{2}} \nu!}{2^{l-\frac{7}{2}} \left(\nu+l+\frac{5}{2}\right)
  \Gamma\left(\nu+l+\frac{1}{2}\right)}\right)^{1/2} \qquad \mbox{\rm (case I)},
\\
  {\cal N}^{(-)}_{\nu} = \left(\frac{\omega^{l+\frac{3}{2}} \nu!}{2^{l-\frac{7}{2}} \left(\nu+l+\frac{3}{2}\right)
  \left(\nu+l+\frac{1}{2}\right) \Gamma\left(\nu+l-\frac{1}{2}\right)}\right)^{1/2} \qquad \mbox{\rm (case II)},
\\
  {\cal N}^{(-)}_{\nu+1} = \left(\frac{\omega^{l+\frac{3}{2}} \nu!\, (\nu+3)}{2^{l-\frac{7}{2}}
  \Gamma\left(\nu+l+\frac{5}{2}\right)}\right)^{1/2} \qquad \mbox{\rm (case III)},
\end{gather*}
and $\nu=0, 1, 2,\ldots$.}

Let us examine in more detail the Laguerre-type polynomials obtained in the f\/irst two cases. From physical considerations related to general properties of SUSYQM, it follows that the wavefunctions $\psi^{(-)}_{\nu}(x)$, $\nu=0, 1, 2,\ldots$, form two complete, orthonormal sets in the Hilbert spaces associated with the corresponding rationally-extended potentials. This suggests that a~similar property should be valid for $\tilde{L}^{(\alpha)}_{1,\nu+2}(z)$ and $\tilde{L}^{(\alpha)}_{2,\nu+2}(z)$ with respect to some appropriate positive-def\/inite measure. From the additional fact that the lowest-degree polynomials are quadratic in $z$ (see Appendix~\ref{appendixA}), we infer that $\tilde{L}^{(\alpha)}_{1,\nu+2}(z)$ and $\tilde{L}^{(\alpha)}_{2,\nu+2}(z)$ are good candidates for the still unknown $X_2$-Laguerre polynomials. Note also that by analogy with classical and $X_1$-Laguerre polynomials, in def\/ining them in (\ref{eq:L-quad-def1}) and (\ref{eq:L-quad-def2}), we have chosen their normalization in such a~way that their highest-degree term is equal to $(-1)^{\nu+2} z^{\nu+2}/\nu!$. Furthermore, it is straightforward to show that the counterparts of equation (\ref{eq:L-lin-comb}) are now linear combinations of f\/ive Laguerre polynomials with constant coef\/f\/icients,
\begin{gather*}
  \tilde{L}^{(\alpha)}_{1,\nu+2}(z)  = (\nu+2)(\nu+1) L^{(\alpha)}_{\nu+2}(z) - 4(\nu+1)(\nu+\alpha+2)
       L^{(\alpha)}_{\nu+1}(z)  \\
\phantom{\tilde{L}^{(\alpha)}_{1,\nu+2}(z)  =}{}  + 2(\nu+\alpha+2)(3\nu+2\alpha+2) L^{(\alpha)}_{\nu}(z) - 4(\nu+\alpha+2)(\nu+\alpha)
       L^{(\alpha)}_{\nu-1}(z)  \\
\phantom{\tilde{L}^{(\alpha)}_{1,\nu+2}(z)  =}{} + (\nu+\alpha+2)(\nu+\alpha-1) L^{(\alpha)}_{\nu-2}(z)
\end{gather*}
and
\begin{gather*}
  \tilde{L}^{(\alpha)}_{2,\nu+2}(z)  = (\nu+2)(\nu+1) L^{(\alpha)}_{\nu+2}(z) - 4(\nu+1)(\nu+\alpha+1)
       L^{(\alpha)}_{\nu+1}(z)  \\
\phantom{\tilde{L}^{(\alpha)}_{2,\nu+2}(z)  =}{}  + 2(\nu+\alpha+1)(3\nu+2\alpha+1) L^{(\alpha)}_{\nu}(z) - 4(\nu+\alpha+1)(\nu+\alpha)
       L^{(\alpha)}_{\nu-1}(z)  \\
\phantom{\tilde{L}^{(\alpha)}_{2,\nu+2}(z)  =}{} + (\nu+\alpha+1)(\nu+\alpha) L^{(\alpha)}_{\nu-2}(z).
\end{gather*}

Since for case III, the wavefunctions $\psi^{(-)}_{\nu+1}(x)$ have to be supplemented with the func\-tion $\phi^{-1}(x)$ in order to form a complete set of orthonormal wavefunctions for the corresponding rationally-extended potential (see equation (\ref{eq:wf-iii})), it is rather clear that the Laguerre-type polynomials $\tilde{L}^{(\alpha)}_{3,\nu+3}(z)$ will not form a complete orthogonal set of exceptional polynomials. One of their properties is however worth pointing out. Although from their def\/inition (\ref{eq:L-quad-def3}) one would expect that their expansion in terms of classical Laguerre polynomials would contain seven terms, one actually gets only f\/ive of them, i.e.,
\begin{gather*}
  \tilde{L}^{(\alpha)}_{3,\nu+3}(z)  = (\nu\!+ 2)(\nu\!+ 1) L^{(\alpha)}_{\nu+3}(z) - 4(\nu\!+ 2)(\nu\!+ 1)
       L^{(\alpha)}_{\nu+2}(z)\! + 2(\nu\!+ 1)(3\nu\!+ \alpha+5) L^{(\alpha)}_{\nu+1}(z)  \\
\phantom{\tilde{L}^{(\alpha)}_{3,\nu+3}(z)  =}{} - 4(\nu+1)(\nu+\alpha+1) L^{(\alpha)}_{\nu}(z) + (\nu+\alpha+1)(\nu+\alpha) L^{(\alpha)}_{\nu-1}(z),
\end{gather*}
the coef\/f\/icients of $L^{(\alpha)}_{\nu-2}(z)$ and $L^{(\alpha)}_{\nu-3}(z)$ vanishing identically. Observe that in (\ref{eq:L-quad-def3}), the polyno\-mials have been normalized in such a way that their highest-degree term is $(-1)^{\nu+3} z^{\nu+3}/[(\nu+3)\nu!]$.

In cases I and II, the extended potential ground-state wavefunction $\psi^{(-)}_0(x)$, expressed in terms of $\tilde{L}^{\left(l+\frac{1}{2}\right)}_{1,2} \left(\frac{1}{2} \omega x^2\right)$ or $\tilde{L}^{\left(l+\frac{1}{2}\right)}_{2,2} \left(\frac{1}{2} \omega x^2\right)$, respectively, can be used to prove that the correspon\-ding potential $V_{l,{\rm ext}}(x)$ is shape invariant, its partner being given by $V_{l+1,{\rm ext}}(x)$ (see Appendix~\ref{appendixD}). This generalizes to the quadratic case a property already demonstrated for the linear one in~\cite{cq08}.

\section{Scarf I potential}\label{section3}

The Scarf I potential \cite{cooper}
\begin{gather*}
    V_{A,B}(x) = [A(A-1) + B^2] \sec^2 x - B(2A-1) \sec x \tan x,
\end{gather*}
def\/ined on the f\/inite interval $- \frac{\pi}{2} < x < \frac{\pi}{2}$, depends on two parameters $A$, $B$, such that $0 < B < A-1$. It has an inf\/inite number of bound-state energies
\begin{gather}
  E^{(A)}_{\nu} = (A + \nu)^2, \qquad \nu = 0, 1, 2, \ldots,  \label{eq:Scarf-E}
\end{gather}
depending only on $A$, and with corresponding wavefunctions
\begin{gather}
\psi^{(A,B)}_{\nu}(x) = {\cal N}^{(A,B)}_{\nu} (1 - \sin x)^{\frac{1}{2}(A-B)} (1 + \sin x)^{\frac{1}{2}(A+B)}
        P^{\left(A-B-\frac{1}{2}, A+B-\frac{1}{2}\right)}_{\nu}(z), \qquad z \equiv \sin x, \nonumber\\
  {\cal N}^{(A,B)}_{\nu} = \left(\frac{(2A+2\nu) \nu!\, \Gamma(2A+\nu)}
       {2^{2A} \Gamma\left(A-B+\nu+\frac{1}{2}\right) \Gamma\left(A+B+\nu+\frac{1}{2}\right)}\right)^{1/2},
 \label{eq:Scarf-wf}
\end{gather}
vanishing at both end points and expressed in terms of Jacobi polynomials.

In the standard SUSY approach corresponding to case {\it i}, deletion of the ground-state energy $E^{(A)}_0$ yields another Scarf I potential $V_{A+1,B}(x)$ (in other words the potential is shape invariant~\cite{gendenshtein}). In such a case, $\phi(x) = (1 - \sin x)^{(A-B)/2} (1 + \sin x)^{(A+B)/2}$ and
\begin{gather*}
  W(x) = A \tan x - B \sec x.
\end{gather*}

As in Section~\ref{section2}, we will now proceed to modify such a superpotential in order to generate rationally-extended Scarf I potentials.

\subsection{Extended classes of rational Scarf I potentials}\label{section3.1}

Let us assume a new superpotential of the form
\begin{gather*}
  W(x) = a \tan x + b \sec x - \frac{dg/dx}{g},
\end{gather*}
where $a$, $b$ are two constants and $g$ is some polynomial in $z = \sin x$. We now demand that $V^{(+)}(x)$ belongs to the Scarf I potential family up to some reparametrization of coef\/f\/icients and that $V^{(-)}(x)$ dif\/fers from $V_{A,B}(x)$ in the presence of some rational terms, i.e.,
\begin{gather*}
  V^{(+)}(x)  = V_{A',B'}(x), \qquad
  V^{(-)}(x)  = V_{A,B,{\rm ext}}(x) = V_{A,B}(x) + V_{A,B,{\rm rat}}(x).
\end{gather*}
Here $V_{A,B,{\rm rat}}(x)$ should be a singularity-free function on the interval $- \frac{\pi}{2} < x < \frac{\pi}{2}$.

\subsubsection{Linear case}

On assuming
\begin{gather*}
  g(\sin x) = \sin x + c, \qquad \frac{dg(\sin x)}{dx} = \cos x, \qquad |c| > 1,
\end{gather*}
and proceeding as in Section~\ref{section2.1.1}, we arrive at the conditions
\begin{gather}
  c = \frac{2b}{2a+1}, \qquad E = (a+1)^2,  \label{eq:Scarf-cond1}
\end{gather}
and
\begin{gather}
  a(a+1) + b^2 = A(A-1) + B^2, \qquad (2a+1) b = - (2A-1) B,  \label{eq:Scarf-cond2}
\end{gather}
coming from $V^{(+)}$ and $V^{(-)}$, respectively.

\sloppy{By successively using equations (\ref{eq:Scarf-cond2}) and (\ref{eq:Scarf-cond1}), we get $(a, b) = \left(\mp B - \frac{1}{2}, \pm A \mp \frac{1}{2}\right)$ or $\left(\mp A - \frac{1}{2} \pm \frac{1}{2}, \pm B\right)$ and $c = - (2A-1)/(2B)$ or $- 2B/(2A-1)$. Since only the former choice for $c$ is compatible with the condition $|c| > 1$, we are only left with $(a, b) = \left(\mp B - \frac{1}{2}, \pm A \mp \frac{1}{2}\right)$, leading to}
\begin{gather}
  V^{(+)}(x) = V_{A, B\pm1}(x),\nonumber
\\
  V^{(-)}(x) = V_{A,B}(x) + \frac{2(2A-1)}{2A-1 - 2B\sin x} - \frac{2[(2A-1)^2 - 4B^2]}{(2A-1 - 2B\sin x)^2}.
  \label{eq:Scarf-pot-lin}
\end{gather}
Correspondingly,
\begin{gather*}
  W(x) = \left(\mp B - \frac{1}{2}\right) \tan x \pm \left(A - \frac{1}{2}\right) \sec x + \frac{2B \cos x}
  {2A-1 - 2B\sin x},
\\
  E = \left(B \mp \tfrac{1}{2}\right)^2,
\\
  \phi(x) = (1 + \sin x)^{\mp\frac{1}{2}\left(A+B-\frac{1}{2}\pm\frac{1}{2}\right)} (1 - \sin x)^{\pm\frac{1}{2}
  \left(A-B-\frac{1}{2}\mp\frac{1}{2}\right)} (2A-1 - 2B \sin x),
\end{gather*}
where we take either all upper or all lower signs.

It is straightforward to see that $E$ and $\phi(x)$ satisfy the conditions for strict isospectrality (case~{\it ii} of SUSYQM). Since, on the other hand, the $B$ value does not inf\/luence the Scarf I energy eigenvalues, it follows that the rationally-extended potential $V^{(-)}(x)$ has the spectrum~(\ref{eq:Scarf-E}). The results are therefore very similar to those of Section~\ref{section2.1.1}: we have two possibilities for the starting reparametrized Scarf I potential, but a single one for the extended potential.

\subsubsection{Quadratic case}\label{section3.1.2}

With the assumption
\begin{gather*}
  g(\sin x) = \sin^2 x + c \sin x + d, \qquad \frac{dg(\sin x)}{dx} = \cos x (2\sin x + c),
\end{gather*}
the discussion gets rather involved. Firstly, one has to impose that the function $g(z)$ has either two real zeros outside the def\/ining interval $-1 < z < 1$ or two complex conjugate ones and, for the former possibility, one has to distinguish between two zeros on the same side of the interval or one zero on the left and the other on the right. As a result, the parameters $c$ and $d$ may vary in three distinct ranges,
\begin{gather*}
  1 < |c| - 1 < d \le \frac{c^2}{4} \qquad \mbox{\rm (case 1a)},
\qquad
  d < - |c| - 1 \qquad \mbox{\rm (case 1b)},
\end{gather*}
and
\begin{gather*}
  d > \frac{c^2}{4} \qquad \mbox{\rm (case 2)}.
\end{gather*}
Secondly, one has to translate these conditions on $c$, $d$ into restrictions on $a$, $b$, taking into account that the two sets of parameters are related to one another through the equations
\begin{gather*}
  c = \frac{4b}{2a+3}, \qquad d = \frac{4b^2 - (2a+3)}{2(a+1)(2a+3)},
\end{gather*}
coming from $V^{(+)}$. Thirdly, the resulting conditions on $a$, $b$ have to be combined with the links $(a, b) = \left(\mp B - \frac{1}{2}, \pm A \mp \frac{1}{2}\right)$ or $\left(\mp A - \frac{1}{2} \pm \frac{1}{2}, \pm B\right)$ imposed by $V^{(-)}$.

This leads to several choices for $(a, b)$ and for the domain of variation of $A$ and $B$, which are summarized in Table~\ref{table1}. Apart from the three options marked with an asterisk (and to be referred to as cases I, II, and III), most of the possibilities are characterized by a very limited range for at least one of the parameters $A$, $B$. For simplicity's sake, they will not be considered any further.

\begin{table}[t]
\centering

\caption{Superpotential parameters for rationally-extended Scarf I potentials in the quadratic case.}\label{table1}
\vspace{1mm}

\begin{tabular}{llll}
  \hline\hline\\[-0.2cm]
  Case & $a$ & $b$ & Conditions\\[0.2cm]
  \hline\\[-0.2cm]
  1a$^*$ & $- B - \frac{1}{2}$ & $A - \frac{1}{2}$ & $1 < B < A-1$ \\[0.2cm]
  1b & $- B - \frac{1}{2}$ & $A - \frac{1}{2}$ & $\frac{3}{2} < A < 2, \quad \frac{1}{2} < B < A-1$ \\[0.2cm]
  1b & $- B - \frac{1}{2}$ & $A - \frac{1}{2}$ & $A \ge 2, \quad \frac{1}{2} < B < 1$ \\[0.2cm]
  2 & $- B - \frac{1}{2}$ & $A - \frac{1}{2}$ & $\frac{5}{4} < A < \frac{3}{2}, \quad \frac{3}{2} - A < B < A-1$
       \\[0.2cm]
  2 & $- B - \frac{1}{2}$ & $A - \frac{1}{2}$ & $A \ge \frac{3}{2}, \quad 0 < B < \frac{1}{2}$ \\[0.2cm]
  2$^*$ & $B - \frac{1}{2}$ & $- A + \frac{1}{2}$ & $0 < B < A - \frac{3}{2}$ \\[0.2cm]
  2 & $-A$ & $B$ & $1 < A \le \frac{5}{4}, \quad 0 < B < A-1$ \\[0.2cm]
  2 & $-A$ & $B$ & $\frac{5}{4} < A < \frac{3}{2}, \quad 0 < B < - A + \frac{3}{2}$ \\[0.2cm]
  2$^*$ & $-A$ & $B$ & $0 < B < A - \frac{3}{2}$ \\[0.2cm]
  \hline \hline
\end{tabular}
\end{table}

For the selected cases, the two partner potentials can be written as
\begin{gather}
  V^{(+)}(x) = V_{A',B'}(x),
\qquad
  V^{(-)}(x) = V_{A,B}(x) + \frac{N_1(x)}{D(x)} + \frac{N_2(x)}{D^2(x)},  \label{eq:Scarf-pot-quad}
\end{gather}
where
\begin{gather}
   A' = A, \qquad B' = B + 1, \nonumber\\
   N_1(x) = - 4 [(2A-1)(2B-1)(2B-2) \sin x + 2(2A-1)^2 - (2B-2)^2 (2B+1)], \nonumber\\
   N_2(x) = - 8(2B-2)(2A-2B+1)(2A+2B-3) \nonumber\\
   \hphantom{N_2(x) =} \times [2(2A-1)(2B-1) \sin x - (2A-1)^2 - 2B (2B-2)],\nonumber \\
   D(x) = (2B-1) [(2B-2) \sin x - (2A-1)]^2 - (2A-2B+1)(2A+2B-3)
  \label{eq:Scarf-pot-I}
\end{gather}
for case I,
\begin{gather}
  A' = A, \qquad B' = B - 1, \nonumber\\
  N_1(x) = - 4 [(2A-1)(2B+1)(2B+2) \sin x - 2(2A-1)^2 - (2B+2)^2 (2B-1)], \nonumber\\
  N_2(x) = 8(2B+2)(2A-2B-3)(2A+2B+1) \nonumber\\
  \hphantom{N_2(x) =} \times [2(2A-1)(2B+1) \sin x - (2A-1)^2 - 2B (2B+2)], \nonumber\\
  D(x) = (2B+1) [(2B+2) \sin x - (2A-1)]^2 + (2A-2B-3)(2A+2B+1)
  \label{eq:Scarf-pot-II}
\end{gather}
for case II, and
\begin{gather}
  A' = A + 1, \qquad B' = B, \nonumber\\
  N_1(x) = - 8 [B(2A-2)(2A-3) \sin x - A(2A-3)^2 + 4B^2], \nonumber\\
  N_2(x) = 8 (2A-3)(2A-2B-3)(2A+2B-3) \nonumber\\
  \hphantom{N_2(x) =}  \times [4B (2A-2) \sin x - 4B^2 - (2A-1)(2A-3)], \nonumber\\
  D(x) = (2A-2) [(2A-3) \sin x - 2B]^2 + (2A-2B-3)(2A+2B-3)
  \label{eq:Scarf-pot-III}
\end{gather}
for case III. Correspondingly,
\begin{gather*}
  E = \left(B - \tfrac{3}{2}\right)^2, \qquad \phi(x) = (1-\sin x)^{\frac{1}{2}(A-B-1)} (1+\sin x)^{-\frac{1}{2}
  (A+B)} D(x),
\\
  E = \left(B + \tfrac{3}{2}\right)^2, \qquad \phi(x) = (1-\sin x)^{-\frac{1}{2}(A-B)} (1+\sin x)^{\frac{1}{2}
  (A+B-1)} D(x),
\end{gather*}
and
\begin{gather*}
  E = (A-2)^2, \qquad \phi(x) = (1-\sin x)^{-\frac{1}{2}(A-B)} (1+\sin x)^{-\frac{1}{2}(A+B)} D(x),
\end{gather*}
from which it follows that cases I and II are characterized by strict isospectrality (case {\it ii} of SUSYQM), whereas in case III, $\phi^{-1}(x)$, which turns out to be normalizable on $\left(- \frac{\pi}{2}, \frac{\pi}{2}\right)$ and to vanish at both end points, is the ground-state wavefunction of $V^{(-)}(x)$ with energy eigenvalue $E^{(-)}_0 = E^{(A+1,B)}_0 - 3(2A-1) = (A-2)^2$ (case {\it iii} of SUSYQM).

We conclude that for Scarf I potential, the quadratic case leads to rather similar results to those obtained for the radial oscillator, the form of the rationally-extended potential becoming sensitive  to the type of reparametrization made for the starting conventional one.

\subsection{Determination of wavefunctions}\label{section3.2}

The procedure used here to determine the wavefunctions of the rationally-extended Scarf I potentials being the same as that introduced in Section~\ref{section2.2}, we will only state the results.

\subsubsection{Linear case}

On choosing for $V^{(+)}(x)$ the Scarf I potential $V_{A,B+1}(x)$, for instance, we obtain for the wavefunctions of $V^{(-)}(x)$
\begin{gather}
  \psi^{(-)}_{\nu}(x) = \frac{{\cal N}^{(+)}_{\nu}}{\sqrt{\varepsilon_{\nu}}} \frac{(1-z)^{\frac{1}{2}\left(\alpha
  + \frac{1}{2}\right)} (1+z)^{\frac{1}{2}\left(\beta + \frac{1}{2}\right)}}{\beta + \alpha - (\beta - \alpha) z}
  \hat{\cal O}^{(\alpha,\beta)}_1 P^{(\alpha-1,\beta+1)}_{\nu}(z),  \label{eq:Scarf-partner-wf}
\end{gather}
where ${\cal N}^{(+)}_{\nu} = {\cal N}^{(A,B-1)}_{\nu}$, $\varepsilon_{\nu} = \left(\nu + A - B + \frac{1}{2}\right)\left(\nu + A + B - \frac{1}{2}\right)$, $\alpha = A - B - \frac{1}{2}$, $\beta = A + B - \frac{1}{2}$, $z = \sin x$, and
\begin{gather*}
  \hat{\cal O}^{(\alpha,\beta)}_1 \equiv [\beta+\alpha - (\beta-\alpha)z] \left((1+z) \frac{d}{dz} + \beta+1
  \right) + (\beta-\alpha)(1+z).
\end{gather*}
The action of this f\/irst-order dif\/ferential operator on the Jacobi polynomial $P^{(\alpha-1,\beta+1)}_{\nu}(z)$ can be easily proved to be given by
\begin{gather*}
  \hat{\cal O}^{(\alpha,\beta)}_1 P^{(\alpha-1,\beta+1)}_{\nu}(z) = 2(\beta-\alpha)(\beta+\nu)
  \hat{P}^{(\alpha,\beta)}_{\nu+1}(z),
\end{gather*}
where $\hat{P}^{(\alpha,\beta)}_{\nu+1}(z)$ is a ($\nu+1$)th-degree polynomial def\/ined by
\begin{gather}
  \hat{P}^{(\alpha,\beta)}_{\nu+1}(z) = - \frac{1}{2} \left(z - \frac{\beta+\alpha}{\beta-\alpha}\right)
  P^{(\alpha,\beta)}_{\nu}(z) \nonumber\\
  \phantom{\hat{P}^{(\alpha,\beta)}_{\nu+1}(z) =}{} + (\beta+\alpha+2\nu)^{-1} \left(\frac{\beta+\alpha}{\beta-\alpha}
  P^{(\alpha,\beta)}_{\nu}(z) - P^{(\alpha,\beta)}_{\nu-1}(z)\right).  \label{eq:P-lin-def}
\end{gather}

The partner potential wavefunctions (\ref{eq:Scarf-partner-wf}) can therefore be written in terms of $\hat{P}^{(\alpha,\beta)}_{\nu+1}(z)$ as
\begin{gather}
  \psi^{(-)}_{\nu}(x) = {\cal N}^{(-)}_{\nu} \frac{(1-\sin x)^{\frac{1}{2}(A-B)} (1+\sin x)^{\frac{1}{2}(A+B)}}
  {2A-1 - 2B \sin x} \hat{P}^{\left(A-B-\frac{1}{2}, A+B-\frac{1}{2}\right)}_{\nu+1}(\sin x),
  \label{eq:Scarf-partner-wf-bis}
\end{gather}
where
\begin{gather}
  {\cal N}^{(-)}_{\nu} = \frac{B}{2^{A-2}} \left(\frac{(2A+2\nu) \nu!\, \Gamma(2A+\nu)}{\left(A{-}B{+}\nu{+}
  \frac{1}{2}\right) \left(A{+}B{+}\nu{+}\frac{1}{2}\right) \Gamma\left(A{-}B{+}\nu{-}\frac{1}{2}\right)
  \Gamma\left(A{+}B{+}\nu{-}\frac{1}{2}\right)}\right)^{1/2}.  \label{eq:Scarf-partner-N}
\end{gather}

Had we started from the Scarf I potential $V_{A,B-1}(x)$, we would have been led to  computing the action of
\begin{gather*}
  \hat{\cal O}^{(\alpha,\beta)}_2 \equiv [\beta+\alpha - (\beta-\alpha)z] \left((1-z) \frac{d}{dz} - (\alpha+1)
  \right) + (\beta-\alpha)(1-z)
\end{gather*}
on $P^{(\alpha+1,\beta-1)}_{\nu}(z)$. Since $\hat{\cal O}^{(\alpha,\beta)}_1$ is changed into $- \hat{\cal O}^{(\alpha,\beta)}_2$ under the permutation of $\alpha$ with $\beta$, combined with the transformation $z \to - z$, we would have obtained the same results~(\ref{eq:Scarf-partner-wf-bis}) and~(\ref{eq:Scarf-partner-N}) as before, up to some irrelevant overall sign.

$\hat{P}^{(\alpha,\beta)}_{\nu+1}(z)$, $\nu=0, 1, 2,\ldots$, form the second complete orthogonal set of polynomials with respect to some positive-def\/inite measure that was constructed in~\cite{gomez08a, gomez08b} by starting with some linear polynomial. As recalled in Appendix~\ref{appendixB}, these $X_1$-Jacobi polynomials can be expressed as linear combinations of three classical Jacobi ones with constant coef\/f\/icients. They are normalized in such a way that their highest-degree term is $- 2^{-\nu-1} \binom{2\nu+\alpha+\beta}{\nu} z^{\nu+1}$ (as compared with $2^{-\nu} \binom{2\nu+\alpha+\beta}{\nu} z^{\nu}$ for $P^{(\alpha,\beta)}_{\nu}(z)$). In the same appendix, two special cases of well-behaved potentials $V^{(-)}(x)$ outside the allowed range $0 < B < A-1$ of parameter values are considered in connection with two so far unknown limiting properties of $\hat{P}^{(\alpha,\beta)}_{\nu+1}(z)$.

Observe that the f\/irst occurrence of the rationally-extended Scarf I potential (\ref{eq:Scarf-pot-lin}) and of the exceptional $X_1$-Jacobi polynomials in quantum mechanics can be traced back to~\cite{cq08}, where it was also demonstrated that such a potential is shape invariant with a partner given by $V_{A+1,B,{\rm ext}}(x)$.

We now plan to generalize $\hat{P}^{(\alpha,\beta)}_{\nu+1}(z)$ to the more sophisticated extended potentials introduced in Section~\ref{section3.1.2}.

\subsubsection{Quadratic case}\label{section3.2.2}

Let us start by considering the rationally-extended potential corresponding to case I and def\/ined in equations (\ref{eq:Scarf-pot-quad}) and (\ref{eq:Scarf-pot-I}). In calculating its wavefunctions, we arrive at the f\/irst-order dif\/ferential operator
\begin{gather}
  \tilde{\cal O}^{(\alpha,\beta)}_1 \equiv {\cal D}(z) \left((1+z) \frac{d}{dz} + \beta+1\right) - (1+z)
  \dot{\cal D}(z),  \label{eq:diff-1}
\end{gather}
where $\alpha = A-B-\frac{1}{2}$, $\beta = A+B-\frac{1}{2}$, and ${\cal D}(z)$ amounts to the function $D(x)$ re-expressed in terms of $z = \sin x$, while $\dot{\cal D}(z)$ denotes its derivative with respect to $z$, i.e.,
\begin{gather*}
  {\cal D}(z) = (\beta\!-\alpha\!-2) [(\beta\!-\alpha\!-1) (\beta\!-\alpha\!-2) z^2\! - 2 (\beta\!-\alpha\!-1) (\beta\!+\alpha) z +
  (\beta\!+\alpha)^2\! + \beta\!-\alpha\!-2],
\\
  \dot{\cal D}(z) = 2 (\beta-\alpha-1) (\beta-\alpha-2) [(\beta-\alpha-2) z - (\beta+\alpha)].
\end{gather*}
Such an operator leads to a new family of ($\nu+2$)th-degree Jacobi-type polynomials $\tilde{P}^{(\alpha,\beta)}_{1,\nu+2}(z)$, def\/ined by
\begin{gather}
    \tilde{\cal O}^{(\alpha,\beta)}_1 P^{(\alpha-1,\beta+1)}_{\nu}(z) = 4 (\beta-\alpha-1) (\beta-\alpha-2)^2
         (\nu+\beta-1) \tilde{P}^{(\alpha,\beta)}_{1,\nu+2}(z) \nonumber\\
    \qquad{}= \frac{\beta-\alpha-2}{2\nu+\beta+\alpha} \Bigl\{\Bigl[(\beta-\alpha-1) (\beta-\alpha-2) (2\nu+\beta+
         \alpha) (\nu+\beta-1) z^2 \nonumber\\
   \qquad \quad {}- 2 (\beta-\alpha-1) (\beta+\alpha) [(\nu+\beta) (2\nu+\beta+\alpha) + \beta-\alpha-2] z \nonumber\\
    \qquad\quad {}+ (\nu+\beta+1) (2\nu+\beta+\alpha) [(\beta+\alpha)^2\! + \beta-\alpha-2] + 2 (\beta-\alpha-1)
         (\beta+\alpha)^2\Bigr] P^{(\alpha,\beta)}_{\nu}(z) \nonumber\\
   \qquad \quad {}+ 4 (\beta-\alpha-1) (\nu+\beta) [(\beta-\alpha-2) z - (\beta+\alpha)] P^{(\alpha,\beta)}_{\nu-1}(z)
         \Bigr\},
  \label{eq:P-quad-def1}
\end{gather}
and the result for $\psi^{(-)}_{\nu}(x)$ reads
\begin{gather*}
  \psi^{(-)}_{\nu}(x) = {\cal N}^{(-)}_{\nu} \frac{(1-\sin x)^{\frac{1}{2}(A-B)} (1+\sin x)^{\frac{1}{2}(A+B)}}
  {D(x)} \tilde{P}^{\left(A-B-\frac{1}{2}, A+B-\frac{1}{2}\right)}_{1,\nu+2}(\sin x),
\end{gather*}
where
\begin{gather*}
  {\cal N}^{(-)}_{\nu}  = \frac{(B-1)^2 (2B-1)}{2^{A-4}} [(2A+2\nu) \nu!\, \Gamma(2A+\nu)]^{1/2} \\
 \phantom{{\cal N}^{(-)}_{\nu}  =}{}  \times \left[\left(A-B+\nu+\tfrac{3}{2}\right) \left(A+B+\nu+\tfrac{1}{2}\right)
        \left(A+B+\nu-\tfrac{1}{2}\right)\right]^{-1/2} \\
\phantom{{\cal N}^{(-)}_{\nu}  =}{} \times \left[\Gamma\left(A-B+\nu-\tfrac{1}{2}\right)
        \Gamma\left(A+B+\nu-\tfrac{3}{2}\right)\right]^{-1/2}.
\end{gather*}

For the case II potential given in equations (\ref{eq:Scarf-pot-quad}) and (\ref{eq:Scarf-pot-II}), it turns out that the f\/irst-order dif\/ferential operator $\tilde{\cal O}^{(\alpha,\beta)}_2$, appearing in the calculation of its wavefunctions, satisf\/ies the same type of symmetry relation with $\tilde{\cal O}^{(\alpha,\beta)}_1$ of equation (\ref{eq:diff-1}) as that connecting $\hat{\cal O}^{(\alpha,\beta)}_2$ with $\hat{\cal O}^{(\alpha,\beta)}_1$ in the linear case. In other words, $\tilde{\cal O}^{(\alpha,\beta)}_2$ can be obtained from $\tilde{\cal O}^{(\alpha,\beta)}_1$ by simultaneously permuting~$\alpha$ with $\beta$ and changing $z$ into $-z$. As a result, no new family of Jacobi-type polynomials arises for such a potential, its wavefunctions being expressed as
\begin{gather*}
  \psi^{(-)}_{\nu}(x) = (-1)^{\nu} {\cal N}^{(-)}_{\nu} \frac{(1-\sin x)^{\frac{1}{2}(A-B)}
  (1+\sin x)^{\frac{1}{2}(A+B)}}{D(x)} \tilde{P}^{\left(A+B-\frac{1}{2}, A-B-\frac{1}{2}\right)}_{1,\nu+2}(-\sin x),
\\
  {\cal N}^{(-)}_{\nu}  = - \frac{(B+1)^2 (2B+1)}{2^{A-4}} [(2A+2\nu) \nu!\, \Gamma(2A+\nu)]^{1/2} \\
\phantom{{\cal N}^{(-)}_{\nu}  =}{}  \times \left[\left(A-B+\nu+\tfrac{1}{2}\right) \left(A-B+\nu-\tfrac{1}{2}\right)
        \left(A+B+\nu+\tfrac{3}{2}\right)\right]^{-1/2} \\
\phantom{{\cal N}^{(-)}_{\nu}  =}{}  \times \left[\Gamma\left(A-B+\nu-\tfrac{3}{2}\right)
        \Gamma\left(A+B+\nu-\tfrac{1}{2}\right)\right]^{-1/2},
\end{gather*}
in terms of the ($\nu+2$)th-degree Jacobi-type polynomials def\/ined in (\ref{eq:P-quad-def1}).

Finally, for the case III potential def\/ined in (\ref{eq:Scarf-pot-quad}) and (\ref{eq:Scarf-pot-III}), the counterpart of (\ref{eq:diff-1}) reads
\begin{gather*}
  \tilde{\cal O}^{(\alpha,\beta)}_3 \equiv {\cal D}(z) \left((1-z^2) \frac{d}{dz} + \beta-\alpha -
  (\beta+\alpha+2)z\right) - (1-z^2) \dot{\cal D}(z),
\end{gather*}
where $\alpha$, $\beta$ are the same as before and
\begin{gather*}
  {\cal D}(z) = (\beta+\alpha-1) [(\beta+\alpha-2) z - (\beta-\alpha)]^2 + (2\alpha-2) (2\beta-2),
\\
  \dot{\cal D}(z) = 2 (\beta+\alpha-1) (\beta+\alpha-2) [(\beta+\alpha-2) z - (\beta-\alpha)].
\end{gather*}
This gives rise to a new family of Jacobi-type polynomials $\tilde{P}^{(\alpha,\beta)}_{3,\nu+3}(z)$, def\/ined by
\begin{gather*}
    \tilde{\cal O}^{(\alpha,\beta)}_3 P^{(\alpha+1,\beta+1)}_{\nu}(z) = 8 (\beta+\alpha-1) (\beta+\alpha-2)^2
         (\beta+\alpha+\nu) \tilde{P}^{(\alpha,\beta)}_{3,\nu+3}(z) \\
  \qquad{} = \frac{1}{(\beta+\alpha+\nu+1) (\beta+\alpha+\nu+2) (\beta+\alpha+2\nu)} \Bigl\{\Bigl[
         (\beta+\alpha+\nu+2) (\beta+\alpha+2\nu+1) \\
  \qquad \quad {}\times [(\beta-\alpha) (\beta+\alpha) - (\beta+\alpha+2\nu) (\beta+\alpha+2\nu+2)z] {\cal D}(z) \\
  \qquad \quad {} + [- (\beta+\alpha+\nu+1) (\beta+\alpha+2\nu) (\beta+\alpha+2\nu+2) \\
  \qquad \quad {} + (\beta-\alpha)^2 \nu + 2 (\beta-\alpha) \nu (\beta+\alpha+2\nu+1) z \\
  \qquad \quad {} + (\beta+\alpha+2\nu) (\beta+\alpha+2\nu+1) (\beta+\alpha+2\nu+2) z^2] \dot{\cal D}(z)\Bigr]
         P^{(\alpha,\beta)}_{\nu}(z) \\
  \qquad \quad {} + 2 (\alpha+\nu) (\beta+\nu) \Bigl[ (\beta+\alpha+\nu+2) (\beta+\alpha+2\nu+2) {\cal D}(z)  \\
  \qquad \quad {} - [(\beta-\alpha) + (\beta+\alpha+2\nu+2) z] \dot{\cal D}(z) \Bigr] P^{(\alpha,\beta)}_{\nu-1}(z)
         \Bigr\}.
\end{gather*}
The extended-potential excited-state wavefunctions can be written as
\begin{gather*}
  \psi^{(-)}_{\nu+1}(x) = {\cal N}^{(-)}_{\nu+1} \frac{(1-\sin x)^{\frac{1}{2}(A-B)} (1+\sin x)^{\frac{1}{2}(A+B)}}
  {D(x)} \tilde{P}^{\left(A-B-\frac{1}{2}, A+B-\frac{1}{2}\right)}_{3,\nu+3}(\sin x),
\\
  {\cal N}^{(-)}_{\nu+1} = \frac{(A-1)(2A-3)^2}{2^{A-3}} \left(\frac{(2A+\nu-1) (2A+2\nu+2) \nu!\,
  \Gamma(2A+\nu+2)}{(\nu+3) \Gamma\left(A-B+\nu+\frac{3}{2}\right)
  \Gamma\left(A+B+\nu+\frac{3}{2}\right)}\right)^{1/2}.
\end{gather*}

It is obvious that the two families of Jacobi-type polynomials that we have just introduced are of a dif\/ferent nature. From physical considerations, the f\/irst set of polynomials $\tilde{P}^{(\alpha,\beta)}_{1,\nu+2}(z)$, whose lowest-degree one is quadratic in $z$ (see Appendix~\ref{appendixC}), is a good candidate for the still unknown complete, orthogonal family of $X_2$-Jacobi polynomials. By analogy with classical and $X_1$-Jacobi polynomials, $\tilde{P}^{(\alpha,\beta)}_{1,\nu+2}(z)$ has been normalized in such a way that its highest-degree term is $2^{-\nu-2} \binom{2\nu+\alpha+\beta}{\nu} z^{\nu+2}$. By contrast, the other family $\tilde{P}^{(\alpha,\beta)}_{3,\nu+3}(z)$, starting with a cubic polynomial, cannot be complete. Note that the highest-degree term is now given by $- 2^{-\nu-3} \binom{2\nu+\alpha+\beta+2}{\nu} z^{\nu+3}$.

Let us emphasize that for Jacobi-type polynomials, no splitting similar to that observed when going from $\hat{L}^{(\alpha)}_{\nu+1}(z)$ to $\tilde{L}^{(\alpha)}_{1,\nu+2}(z)$ and $\tilde{L}^{(\alpha)}_{2,\nu+2}(z)$ has been encountered.

Finally, in cases I and II, the polynomial $\tilde{P}^{\left(A-B-\frac{1}{2}, A+B-\frac{1}{2}\right)}_{1,2}(\sin x)$ (or its counterpart) can be inserted in the extended potential ground-state wavefunction $\psi^{(-)}_0(x)$ to prove that the corresponding potential $V_{A,B,{\rm ext}}(x)$ is shape invariant, its partner being given by $V_{A+1,B,{\rm ext}}(x)$ (see Appendix~\ref{appendixD}). This again generalizes a result of~\cite{cq08} to the quadratic case.

\section{Final comments}\label{section4}

In the present paper, we have generated new exactly solvable rationally-extended radial oscillator and Scarf I potentials and we have constructed their bound-state wavefunctions. This has been made possible by generalizing  a constructive SUSYQM method recently proposed in~\cite{bagchi08} and based on some reparametrization of the conventional superpotential, together with the addition of a rational term expressed in terms of a polynomial $g(z)$, where $z$ is some appropriately chosen function of $x$. The cases of linear and quadratic polynomials have been considered here.

In the linear case, there appears a single rationally-extended potential, but it can be obtained by starting from an isospectral conventional potential with two distinct sets of reparametrized couplings. In contrast, the quadratic case leads to a variety of rationally-extended potentials, each of them being the partner of a single reparametrized conventional potential. Some potential pairs turn out to be isospectral, while in others, the rational potential has an extra bound state below the conventional potential spectrum.

All rationally-extended potentials belonging to isospectral pairs have been demonstrated to be shape invariant as their conventional counterparts.

Furthermore, by applying our SUSYQM approach, we have explicitly shown that the ($\nu+1$)th-degree polynomials  ($\nu=0, 1, 2,\ldots$) occurring in the bound-state wavefunctions of the extended potentials corresponding to the linear case are the $X_1$-Laguerre or $X_1$-Jacobi polynomials that were recently proved to form two sets of complete, orthogonal exceptional polynomials. We have then proposed several extensions of these polynomials valid for the quadratic case. Among them, we have identif\/ied two dif\/ferent kinds of ($\nu+2$)th-degree Laguerre-type polynomials and a single one of ($\nu+2$)th-degree Jacobi-type polynomials, which are candidates for the still unknown $X_2$-Laguerre and $X_2$-Jacobi exceptional polynomials, respectively.

Two interesting properties, dealt with in~\cite{bagchi08}, have not been considered here, but are worth mentioning. To begin with, our f\/irst-order SUSY transformation may be combined with another one relating conventional potentials with dif\/ferent parameters to produce a reducible second-order SUSY transformation connecting conventional and extended potentials with the same parameters. In the linear case, this results in a second-order transformation admitting two distinct factorizations.

The other point has to do with the possibility of discarding the restriction to real potentials, which has been implicitly made here. Considering also $\cal PT$-symmetric complex potentials indeed facilitates reconciling our approach to the requirement that the rationally-extended potentials be singularity free, hence generating so far unknown complex potentials with a real spectrum. The generalization of the present work along these lines is therefore an interesting topic for future investigation.

When the present work was in its f\/inal stage, there appeared two interesting preprints, whose subjects are related to those considered here. In the f\/irst one \cite{bagchi09}, the existence of distinct factorizations of second-order SUSY transformations into products of two f\/irst-order ones, observed for the rationally-extended generalized P\"oschl--Teller potential of~\cite{bagchi08} and which would also be obtained here in the linear case, is discussed in the framework of type A 2-fold SUSY~\cite{aoyama}. Necessary and suf\/f\/icient conditions for such a situation to occur are derived and some relations to second-order parasupersymmetry and generalized 2-fold superalgebras are noted.

In the second work \cite{odake}, three inf\/inite families of shape-invariant, rationally-extended radial oscillator, trigonometric and hyperbolic P\"oschl--Teller potentials are presented. They are obtained by deforming the corresponding conventional potentials in terms of their degree $\ell$ polynomial eigenfunctions and their bound-state wavefunctions are expressed in terms of Laguerre-type or Jacobi-type polynomials. In the radial oscillator and trigonometric P\"oschl--Teller cases, the f\/irst member of the inf\/inite family, corresponding to $\ell = 1$, is shown to coincide with one of the potentials introduced in~\cite{cq08} and re-obtained here in the linear case for the radial oscillator and Scarf I, respectively.

The general expressions provided both for the potentials and the polynomials in~\cite{odake} enable us to pursue the comparison with the results of~\cite{bagchi08} and those derived here. To start with, it can be easily seen that the f\/irst member (with $\ell = 1$) of the third family (hyperbolic P\"oschl--Teller) actually corresponds via some changes of variable and of parameters ($x \to x/2$, $g \to B-A-1$, $h \to B+A+1$) to the extended generalized P\"oschl--Teller potential constructed in~\cite{bagchi08}. Furthermore, the results corresponding to the second member (with $\ell=2$) of the f\/irst two families agree with those of the present paper associated with the quadratic case and referred to as case I extended radial oscillator and case~II extended Scarf~I, respectively. It should be stressed that no case II extended radial oscillator (with
its corresponding Laguerre-type polynomials) is found there. Whether the existence of such an alternative branch of potentials and polynomials, demonstrated in the quadratic case in the present paper, could be generalized to higher-degree polynomials $g(z)$ would be an interesting topic for future investigation.

\appendix

\section{Examples of Laguerre-type polynomials}\label{appendixA}

In this appendix, we list the f\/irst few $\tilde{L}^{(\alpha)}_{1,\nu+2}$, $\tilde{L}^{(\alpha)}_{2,\nu+2}$, and $\tilde{L}^{(\alpha)}_{3,\nu+3}$ Laguerre-type polynomials and, for comparison's sake, also the corresponding classical and $X_1$-Laguerre polynomials.

\medskip

\noindent
{\sl Laguerre polynomials}
\begin{gather*}
  L^{(\alpha)}_0(z)   = 1, \\
  L^{(\alpha)}_1(z)   = - z + \alpha+1,  \\
  L^{(\alpha)}_2(z)   = \tfrac{1}{2} [z^2 - 2 (\alpha+2) z + (\alpha+2) (\alpha+1)].
\end{gather*}
{\sl $X_1$-Laguerre polynomials}
\begin{gather*}
  \hat{L}^{(\alpha)}_1(z)  = - z - \alpha-1,  \\
  \hat{L}^{(\alpha)}_2(z)  = z^2 - \alpha (\alpha+2),  \\
  \hat{L}^{(\alpha)}_3(z)  = \tfrac{1}{2} [- z^3 + (\alpha+3) z^2 + \alpha (\alpha+3) z  - \alpha (\alpha+1)
          (\alpha+3)].
\end{gather*}
{\sl New Laguerre-type polynomials}
\begin{gather*}
  \tilde{L}^{(\alpha)}_{1,2}(z)  = z^2 + 2 (\alpha+2) z + (\alpha+2) (\alpha+1), \\
  \tilde{L}^{(\alpha)}_{1,3}(z)  = - z^3 - (\alpha+3) z^2 + \alpha (\alpha+3) z + \alpha (\alpha+1) (\alpha+3), \\
  \tilde{L}^{(\alpha)}_{1,4}(z)  = \tfrac{1}{2} [z^4 - 2 (\alpha+1) (\alpha+4) z^2 + \alpha (\alpha+1)^2
         (\alpha+4)],
\\
  \tilde{L}^{(\alpha)}_{2,2}(z)  = z^2 + 2 \alpha z + \alpha (\alpha+1), \\
  \tilde{L}^{(\alpha)}_{2,3}(z)  = - z^3 - (\alpha-1) z^2 + (\alpha+2) (\alpha-1) z + (\alpha+2) (\alpha+1)
         (\alpha-1), \\
  \tilde{L}^{(\alpha)}_{2,4}(z)  = \tfrac{1}{2} [z^4 - 4 z^3 - 2 (\alpha+3) (\alpha-1) z^2 + (\alpha+3)
         (\alpha+2) \alpha (\alpha-1)],
\\
  \tilde{L}^{(\alpha)}_{3,3}(z)  = \tfrac{1}{3} [- z^3 + 3 (\alpha-1) z^2 - 3 \alpha (\alpha-1) z + (\alpha+1)
        \alpha (\alpha-1)], \\
  \tilde{L}^{(\alpha)}_{3,4}(z)  = \tfrac{1}{4} [z^4 - 4 \alpha z^3 + 2 (\alpha-1) (3\alpha+4) z^2 - 4 (\alpha+2)
        \alpha (\alpha-1) z \\
 \phantom{\tilde{L}^{(\alpha)}_{3,4}(z)  =}{}  + (\alpha+2) (\alpha+1) \alpha (\alpha-1)], \\
  \tilde{L}^{(\alpha)}_{3,5}(z)  = \tfrac{1}{10} [- z^5 + 5 (\alpha+1) z^4 - 10 (\alpha^2+2\alpha-1) z^3 + 10
        (\alpha+3) (\alpha+1) (\alpha-1) z^2 \\
 \phantom{\tilde{L}^{(\alpha)}_{3,5}(z) =}{}  - 5 (\alpha+3) (\alpha+2) \alpha (\alpha-1) z + (\alpha+3) (\alpha+2) (\alpha+1) \alpha (\alpha-1)].
\end{gather*}

\section[Limiting cases of extended Scarf I potentials and of $X_1$-Jacobi polynomials]{Limiting cases of extended Scarf I potentials\\ and of $\boldsymbol{X_1}$-Jacobi polynomials}\label{appendixB}

The purpose of this appendix is to review two cases where although the parameter $B$ in the rationally-extended Scarf I potential (\ref{eq:Scarf-pot-lin}) takes a value outside the allowed range $0 < B < A-1$, the potential remains physically acceptable and reduces to some known conventional potential. As a result, there exists a limiting relation between the wavefunctions of the former, given in equations (\ref{eq:Scarf-partner-wf-bis}) and (\ref{eq:Scarf-partner-N}), and those of the latter, expressed in terms of some classical polynomials. The corresponding properties of the $X_1$-Jacobi polynomials will be demonstrated by starting from their known ones, proved in~\cite{gomez08b}.

If we set $B=0$ in equation (\ref{eq:Scarf-pot-lin}), the sum of the two rational terms vanishes and we get the $B \to 0$ limit of the Scarf I potential, which is the one-parameter trigonometric P\"oschl--Teller potential $V_{A,0}(x) = A(A-1) \sec^2 x$. Its energy spectrum is given by equation (\ref{eq:Scarf-E}) and the corresponding wavefunctions can be expressed in terms of Gegenbauer polynomials as \cite{cq99}
\begin{gather*}
  \psi^{(A,0)}_{\nu}(x) = \bar{\cal N}^{(A)}_{\nu} (\cos x)^A C^{(A)}_{\nu}(\sin x),
\end{gather*}
where
\begin{gather*}
  \bar{\cal N}^{(A)}_{\nu} = \left(\frac{\Gamma(A) \Gamma(2A) \nu!\, (A+\nu)}{\sqrt{\pi}\, \Gamma\left(A+
  \frac{1}{2}\right) \Gamma(2A+\nu)}\right)^{1/2}.
\end{gather*}
Comparison with equations (\ref{eq:Scarf-partner-wf-bis}) and (\ref{eq:Scarf-partner-N}) leads to the relation
\begin{gather}
  \lim_{\beta \to \alpha} (\beta-\alpha) \hat{P}^{(\alpha,\beta)}_{\nu+1}(z) = \frac{(\alpha+\nu+1)
  \Gamma(2\alpha+1) \Gamma(\alpha+\nu)}{\Gamma(\alpha) \Gamma(2\alpha+\nu+1)}
  C^{\left(\alpha+\frac{1}{2}\right)}_{\nu}(z).  \label{eq:limit-1}
\end{gather}

The direct proof  of equation (\ref{eq:limit-1}) is based on the expansion of $X_1$-Jacobi polynomials as linear combinations  of three classical Jacobi ones,
\begin{gather*}
  \hat{P}^{(\alpha,\beta)}_{\nu+1}(z)  = - \frac{(\nu+1) (\beta+\alpha+\nu+1)}{(\beta+\alpha+2\nu+1)
       (\beta+\alpha+2\nu+2)} P^{(\alpha,\beta)}_{\nu+1}(z) \\
\phantom{\hat{P}^{(\alpha,\beta)}_{\nu+1}(z)  =}{}  + 2 \frac{\beta+\alpha}{\beta-\alpha} \frac{(\alpha+\nu+1) (\beta+\nu+1)}{(\beta+\alpha+2\nu)
       (\beta+\alpha+2\nu+2)} P^{(\alpha,\beta)}_{\nu}(z)  \\
\phantom{\hat{P}^{(\alpha,\beta)}_{\nu+1}(z)  =}{} - \frac{(\alpha+\nu+1) (\beta+\nu+1)}{(\beta+\alpha+2\nu) (\beta+\alpha+2\nu+1)}
       P^{(\alpha,\beta)}_{\nu-1}(z).
\end{gather*}
On letting $\beta$ go to $\alpha$ in $(\beta-\alpha) \hat{P}^{(\alpha,\beta)}_{\nu+1}(z)$, it is clear that the only surviving term is proportional to~$P^{(\alpha,\alpha)}_{\nu}(z)$, which is known to be expressible in terms of $C^{\left(\alpha+\frac{1}{2}\right)}_{\nu}(z)$ \cite{abramowitz}, thus leading to equation~(\ref{eq:limit-1}).

On the other hand, we observe that although the Scarf I potential $V_{A,B}(x)$ is not def\/ined for $B = A - \frac{1}{2}$ or $A = B + \frac{1}{2}$, the same is not true for its extension $V_{A,B,{\rm ext}}(x)$, given in equation (\ref{eq:Scarf-pot-lin}). The latter is indeed equivalent to the well-behaved conventional Scarf I potential $V_{A+1,A-\frac{3}{2}}(x)$ with energy spectrum $E^{(A+1)}_{\nu} = (A+\nu+1)^2$, $\nu=0, 1, 2,\ldots$, and wavefunctions $\psi^{\left(A+1,A-\frac{3}{2}\right)}_{\nu}(x)$, obtainable from equation (\ref{eq:Scarf-wf}). For the same parameter values, we get from equation (\ref{eq:Scarf-partner-wf-bis})
\begin{gather*}
  \lim_{B \to A-\frac{1}{2}} \psi^{(-)}_0(x) \propto (1-\sin x)^{- \frac{3}{4}} (1+\sin x)^{A-\frac{1}{4}}
  [2A+1 - (2A-1) \sin x],
\end{gather*}
which does not vanish for $x \to \frac{\pi}{2}$, hence is not physically acceptable. This explains the absence of an eigenvalue $A^2$ in the energy spectrum. The presence of the remaining eigenvalues, corresponding to $\nu=1, 2,\ldots$, in (\ref{eq:Scarf-E}) hints at a limiting relation between $\hat{P}^{\left(A-B-\frac{1}{2}, A+B-\frac{1}{2}\right)}_{\nu+1}(z)$ and $P^{(2,2A-1)}_{\nu-1}(z)$ when $B \to A-\frac{1}{2}$. In terms of $\alpha$ and $\beta$, such a relation can be written as
\begin{gather}
  \lim_{\alpha \to 0} \hat{P}^{(\alpha,\beta)}_{\nu+1}(z) = - \frac{\beta+\nu+1}{4\nu} (1-z)^2
  P^{(2,\beta)}_{\nu-1}(z), \qquad \nu = 1, 2, \ldots.  \label{eq:limit-2}
\end{gather}

To prove equation (\ref{eq:limit-2}), let us start from the $\alpha \to 0$ limit of the def\/ining relation (\ref{eq:P-lin-def}) of $X_1$-Jacobi polynomials,
\begin{gather*}
  \lim_{\alpha \to 0} \hat{P}^{(\alpha,\beta)}_{\nu+1}(z) = - \frac{1}{2} (z-1) P^{(0,\beta)}_{\nu}(z) +
  \frac{P^{(0,\beta)}_{\nu}(z) - P^{(0,\beta)}_{\nu-1}(z)}{\beta+2\nu}.
\end{gather*}
On using equations (22.7.18) and (22.7.15) of~\cite{abramowitz}, it is straightforward to transform the latter into
\begin{gather*}
  \lim_{\alpha \to 0} \hat{P}^{(\alpha,\beta)}_{\nu+1}(z) = \frac{(\beta+\nu+1)}{2\nu (\beta+2\nu+1)} (1-z)
  \left[\nu P^{(1,\beta)}_{\nu}(z) - (\nu+1) P^{(1,\beta)}_{\nu-1}(z)\right].
\end{gather*}
Another application of equation (22.7.15) then leads to the searched for result (\ref{eq:limit-2}).

\section{Examples of Jacobi-type polynomials}\label{appendixC}

In this appendix, we list the f\/irst few $\tilde{P}^{(\alpha,\beta)}_{1,\nu+2}$ and $\tilde{P}^{(\alpha,\beta)}_{3,\nu+3}$ Jacobi-type polynomials and, for comparison's sake, also the corresponding classical and $X_1$-Jacobi polynomials.

\medskip

\noindent
{\sl Jacobi polynomials}
\begin{gather*}
  P^{(\alpha,\beta)}_0(z)   = 1, \\
  P^{(\alpha,\beta)}_1(z) , = \tfrac{1}{2} [(\beta+\alpha+2) z - (\beta-\alpha)].
\end{gather*}
{\sl $X_1$-Jacobi polynomials}
\begin{gather*}
  \hat{P}^{(\alpha,\beta)}_1(z)   = \frac{1}{2(\beta-\alpha)} [- (\beta-\alpha) z + \beta+\alpha+2],  \\
  \hat{P}^{(\alpha,\beta)}_2(z)   = \frac{1}{4(\beta-\alpha)} \{- (\beta-\alpha) (\beta+\alpha+2) z^2 +
       [(\beta-\alpha)^2 + (\beta+\alpha) (\beta+\alpha+4)] z \\
\phantom{\hat{P}^{(\alpha,\beta)}_2(z)=}{} - (\beta-\alpha) (\beta+\alpha+2)\}.
  \end{gather*}
{\sl New Jacobi-type polynomials}
\begin{gather*}
  \tilde{P}^{(\alpha,\beta)}_{1,2}(z)   = \frac{1}{4 (\beta-\alpha-1) (\beta-\alpha-2)} [(\beta-\alpha-1)
        (\beta-\alpha-2) z^2 \\
   \phantom{\tilde{P}^{(\alpha,\beta)}_{1,2}(z)   =}{}  - 2 (\beta-\alpha-1) (\beta+\alpha+2) z + (\beta+\alpha+2)^2 + \beta-\alpha-2], \\
  \tilde{P}^{(\alpha,\beta)}_{1,3}(z)   = \frac{1}{8 (\beta-\alpha-1) (\beta-\alpha-2)} \{(\beta-\alpha-1)
        (\beta-\alpha-2) (\beta+\alpha+2) z^3 \\
\phantom{\tilde{P}^{(\alpha,\beta)}_{1,3}(z)   =}{} - (\beta-\alpha-1) [(\beta-\alpha) (\beta-\alpha-2) + 2 (\beta+\alpha) (\beta+\alpha+4)] z^2 \\
\phantom{\tilde{P}^{(\alpha,\beta)}_{1,3}(z)   =}{} + (\beta+\alpha+2) [(\beta-\alpha-2) (2\beta-2\alpha+3) + (\beta+\alpha) (\beta+\alpha+4)] z \\
\phantom{\tilde{P}^{(\alpha,\beta)}_{1,3}(z)   =}{} - (\beta-\alpha) (\beta+\alpha+2)^2 - (\beta-\alpha-4) (\beta-\alpha+2)\},
\\
  \tilde{P}^{(\alpha,\beta)}_{3,3}(z)   = \frac{1}{8 (\beta+\alpha) (\beta+\alpha-1) (\beta+\alpha-2)} \{- (\beta
        +\alpha) (\beta+\alpha-1) (\beta+\alpha-2) z^3 \\
\phantom{\tilde{P}^{(\alpha,\beta)}_{3,3}(z)   =}{}  + 3 (\beta+\alpha) (\beta+\alpha-1) (\beta-\alpha) z^2 - 3 (\beta+\alpha) [(\beta-\alpha)^2 + \beta
        +\alpha-2] z \\
\phantom{\tilde{P}^{(\alpha,\beta)}_{3,3}(z)   =}{} + (\beta-\alpha) [(\beta-\alpha)^2 + 3\beta+3\alpha-4]\}, \\
  \tilde{P}^{(\alpha,\beta)}_{3,4}(z)   = \frac{1}{16 (\beta+\alpha-1) (\beta+\alpha-2) (\beta+\alpha+1)} \\
\phantom{\tilde{P}^{(\alpha,\beta)}_{3,4}(z)   =}{}  \times \{- (\beta+\alpha-2) (\beta+\alpha-1) (\beta+\alpha+1) (\beta+\alpha+4) z^4 \\
\phantom{\tilde{P}^{(\alpha,\beta)}_{3,4}(z)   =}{} + 4 (\beta-\alpha) (\beta+\alpha-1) (\beta+\alpha+1) (\beta+\alpha+2) z^3 \\
\phantom{\tilde{P}^{(\alpha,\beta)}_{3,4}(z)   =}{} - 2 (\beta+\alpha+1) [(\beta-\alpha)^2 (3\beta+3\alpha+2) + (\beta+\alpha-2) (\beta+\alpha+4)] z^2
        \\
\phantom{\tilde{P}^{(\alpha,\beta)}_{3,4}(z)   =}{} + 4 (\beta-\alpha) (\beta+\alpha+1) [(\beta-\alpha)^2 + \beta+\alpha-2] z \\
\phantom{\tilde{P}^{(\alpha,\beta)}_{3,4}(z)   =}{} - (\beta-\alpha)^4 - 2 (\beta-\alpha)^2 (\beta+\alpha-4) + (\beta+\alpha-2) (\beta+\alpha+4)\}.
\end{gather*}

\section{Shape invariance of rationally-extended potentials}\label{appendixD}

In this appendix, we prove the shape invariance of the rationally-extended radial oscillator potentials, def\/ined in equations (\ref{eq:RHO-pot-I}) and (\ref{eq:RHO-pot-II-III}) (the latter with upper signs only), as well as that of the rationally-extended Scarf I potentials given in equations (\ref{eq:Scarf-pot-quad}), (\ref{eq:Scarf-pot-I}), and (\ref{eq:Scarf-pot-II}).

In such a picture corresponding to case {\it i} of SUSYQM, the rationally-extended potential is considered as the starting potential $\tilde{V}^{(+)}(x)$ and its partner is determined from equation~(\ref{eq:V+/-}) as $\tilde{V}^{(-)}(x) = \tilde{V}^{(+)}(x) + 2 \tilde{W}'(x)$, where the superpotential is now given by $\tilde{W}(x) = - d[\ln \tilde{\psi}^{(+)}_0(x)]/dx$, with $\tilde{\psi}^{(+)}_0(x) = \psi^{(-)}_0(x)$.

On using the expressions found for $\psi^{(-)}_0(x)$ in Section~\ref{section2.2.2} and the results of Appendix~\ref{appendixA}, it is straightforward to show that for case I and II extended radial oscillator potentials,
\begin{gather*}
  \tilde{W}(x) = \tilde{W}_1(x) + \tilde{W}_2(x), \qquad \tilde{W}_1(x) = \frac{1}{2} \omega x - \frac{l+1}{x},
\end{gather*}
and
\begin{gather*}
  \tilde{W}_2(x) = 4 \omega x \left(\frac{\omega x^2 + 2l +3}{(\omega x^2 + 2l +3)^2 - 2(2l+3)} -
  \frac{\omega x^2 + 2l +5}{(\omega x^2 + 2l +5)^2 - 2(2l+5)} \right) \qquad ({\rm case\ I}), \\
  \tilde{W}_2(x) = 4 \omega x \left(\frac{\omega x^2 + 2l -1}{(\omega x^2 + 2l -1)^2 + 2(2l-1)} -
  \frac{\omega x^2 + 2l +1}{(\omega x^2 + 2l +1)^2 + 2(2l+1)} \right) \qquad ({\rm case\ II}),
\end{gather*}
from which it directly follows that
\begin{gather*}
  2 \tilde{W}'(x) = - V_{l,{\rm ext}}(x) + V_{l+1,{\rm ext}}(x) + \omega.
\end{gather*}
Hence, in such cases, the partner of $V_{l,{\rm ext}}(x)$ is $V_{l+1,{\rm ext}}(x) + \omega$, which proves the shape invariance of the former.

Similarly, from Section~\ref{section3.2.2} and Appendix~\ref{appendixC}, we obtain for case I and II extended Scarf I potentials,
\begin{gather*}
  \tilde{W}(x) = \tilde{W}_1(x) + \tilde{W}_2(x), \qquad \tilde{W}_1(x) = A \tan x - B \sec x,
\end{gather*}
and
\begin{gather*}
  \tilde{W}_2(x)  = 2 (2B-1)(2B-2) \cos x \\
 \phantom{\tilde{W}_2(x)  =}{}  \times \left(\frac{(2B-2) \sin x - (2A-1)}{D_{A,B}(x)} -
        \frac{(2B-2) \sin x - (2A+1)}{D_{A+1,B}(x)} \right) \qquad ({\rm case\ I}), \\
  \tilde{W}_2(x)  = 2 (2B+1)(2B+2) \cos x \\
\phantom{\tilde{W}_2(x)  =}{} \times \left(\frac{(2B+2) \sin x - (2A-1)}{D_{A,B}(x)} -
        \frac{(2B+2) \sin x - (2A+1)}{D_{A+1,B}(x)} \right) \qquad ({\rm case\ II}),
\end{gather*}
where $D_{A,B}(x)$ denotes the denominator function $D(x)$ of equations (\ref{eq:Scarf-pot-I}) and (\ref{eq:Scarf-pot-II}), associated with some specif\/ied parameters $A$, $B$. Since
\begin{gather*}
  2 \tilde{W}'(x) = - V_{A,B,{\rm ext}}(x) + V_{A+1,B,{\rm ext}}(x),
\end{gather*}
the partner of $V_{A,B,{\rm ext}}(x)$ is $V_{A+1,B,{\rm ext}}(x)$, thus completing the proof.

\subsection*{Acknowledgements}

The author would like to thank B.~Bagchi and R.~Roychoudhury for some interesting discussions.

\newpage

\pdfbookmark[1]{References}{ref}
\LastPageEnding

\end{document}